\documentclass[lettersize,journal]{IEEEtran}
\usepackage{amsmath,amsfonts}
\usepackage{array}
\usepackage[caption=false,font=normalsize,labelfont=sf,textfont=sf]{subfig}
\usepackage{textcomp}
\usepackage{stfloats}
\usepackage{pifont}
\usepackage{url}
\usepackage{verbatim}
\usepackage{graphicx}
\def\BibTeX{{\rm B\kern-.05em{\sc i\kern-.025em b}\kern-.08em
    T\kern-.1667em\lower.7ex\hbox{E}\kern-.125emX}}
\usepackage{balance}
\usepackage{float}

\usepackage{amsmath,amssymb,amsfonts}
\usepackage{xr-hyper}
\usepackage{hyperref}
\usepackage{graphicx}
\usepackage{textcomp}
\usepackage{array}
\usepackage{xcolor}
\usepackage{svg}
\usepackage{amsmath}
\usepackage[linesnumbered,ruled,vlined]{algorithm2e}
\usepackage{algpseudocode}
\usepackage{multirow}
\usepackage{wrapfig}
\usepackage{booktabs}
\usepackage{threeparttable}
\usepackage{caption}
\usepackage[most]{tcolorbox}

\usepackage{cite}
\usepackage{hyperref}

\def\bf{\textbf}

\newtcolorbox[auto counter]{custombox}[1][]{
  enhanced,
  breakable,
  title={Listing \thetcbcounter},
  boxrule=0.5mm,  % Border thickness
  left=0.5mm,      % No left padding
  right=0.5mm,     % No right padding
  top=0.5mm,       % No top padding
  bottom=0.5mm,    % No bottom padding
  #1
}

\newtcolorbox{summarybox}{
  enhanced,
  boxrule=0.3mm,  % Border thickness
  left=0.5mm,      % No left padding
  right=0.5mm,     % No right padding
  top=0.5mm,       % No top padding
  bottom=0.5mm,    % No bottom padding
}

\usepackage{subcaption}
\usepackage{enumerate}
\usepackage{caption}
\usepackage{listings}
\usepackage{enumitem}

\newcommand{\rev}[1]{\textcolor{black}{#1}}
\newcommand{\revtwo}[1]{\textcolor{black}{#1}}
\newcommand{\revth}[1]{\textcolor{black}{#1}}
\newcommand{\revm}[1]{\textcolor{black}{#1}}
\usepackage{balance}
\def\BibTeX{{\rm B\kern-.05em{\sc i\kern-.025em b}\kern-.08em
    T\kern-.1667em\lower.7ex\hbox{E}\kern-.125emX}}

\begin{document}

\title{TriagerX: Dual Transformers for Bug Triaging Tasks with Content and Interaction Based Rankings}

\author{
    Md Afif Al Mamun$^1$, Gias Uddin$^2$, Lan Xia$^3$, Longyu Zhang$^4$ \\ 
    $^1$ University of Calgary, $^2$ York University, $^{3,4}$ IBM Canada
}
% \author{
%     Md Afif Al Mamun$^1$\thanks{University of Calgary, afif.mamun@ucalgary.ca}, 
%     Gias Uddin$^2$\thanks{York University, guddin@yorku.ca}, 
%     Lan Xia$^3$\thanks{IBM Canada, lan\_xia@ca.ibm.com}, 
%     Longyu Zhang$^4$\thanks{IBM Canada, longyu.zhang@ibm.com}
% }

\maketitle

\begin{abstract}
Pretrained Language Models or PLMs are transformer-based architectures that can be used in bug triaging tasks. PLMs can better capture token semantics than traditional Machine Learning (ML) models that rely on statistical features (e.g., TF-IDF, bag of words). However, PLMs may still attend to less relevant tokens in a bug report, which can impact their effectiveness. In addition, the model can be sub-optimal with its recommendations when the interaction history of developers around similar bugs is not taken into account. We designed TriagerX to address these limitations. First, to assess token semantics more reliably, we leverage a dual-transformer architecture. Unlike current state-of-the-art (SOTA) baselines that employ a single transformer architecture, TriagerX collects recommendations from two transformers with each offering recommendations via its last three layers. This setup generates a robust content-based ranking of candidate developers. TriagerX then refines this ranking by employing a novel interaction-based ranking methodology, which considers developers’ historical interactions with similar fixed bugs. Across five datasets, TriagerX surpasses all nine transformer-based methods, including SOTA baselines, often improving Top-1 and Top-3 developer recommendation accuracy by over 10\%. We worked with our large industry partner to successfully deploy TriagerX in their development environment. 
The partner required both developer and component recommendations, with components acting as proxies for team assignments—particularly useful in cases of developer turnover or team changes. We trained TriagerX on the partner’s dataset for both tasks, and it outperformed SOTA baselines by up to 10\% for component recommendations and 54\% for developer recommendations. Replication package. \url{https://github.com/afifaniks/triagerX}
\end{abstract}


\maketitle

\section{Introduction}
\label{sec:intro}

Bug triaging involves assigning reported issues to the most suitable developer or software team for resolution. Over the past few decades, various information retrieval (IR), machine learning (ML), and deep learning (DL) approaches automated this process \cite{Cubranic2004AutomaticBT, anvik2006svm, anvik2011reducing, lee2017dlbasedtriager, mani2018deeptriage, zaidi2020cnntriage, lee2023lbtp}. However, their real-world adoption remains limited due to inconsistent performance across different datasets and industrial settings \cite{sarkar2019ericsson}.

To understand and address these challenges, in collaboration with our industrial partner (IBM), we examined the limitations of existing approaches and then designed a novel bug triaging technique called TriagerX. We have successfully deployed TriagerX within the partner’s development environment. 
%, collecting feedback for further improvements and evaluating its effectiveness in real-world workflows.

TriagerX is built on the transformer architecture of Pretrained Language Models (PLMs). 
This design decision is motivated by recent studies that demonstrated that PLMs outperform traditional ML and Information Retrieval (IR) approaches in large-scale bug assignment tasks, benefiting from context-aware embeddings \cite{zaidi2020cnntriage, lee2023lbtp, dipongkar2023comparison, mani2018deeptriage}. Traditional TF-IDF-based methods rely on lexical similarity and are often less effective for natural language processing tasks where contextual meanings are important \cite{devlin2018bert, peters-etal-2018-deep}. 

Compared to the recent PLM-based bug triaging techniques, we innovate in TriagerX by offering a dual transformer architecture. This architecture decision is grounded on our studies on the effectiveness of current PLMs in the industrial setting. We fine-tuned top-performing PLMs identified in a recent comparative study \cite{dipongkar2023comparison} and compared them with TF-IDF, the state-of-the-art (SOTA) bug triaging model LBT-P (which uses a context-sensitive PLM) \cite{lee2023lbtp}, and DBRNN-A \cite{mani2018deeptriage}, which relies on context-insensitive Word2Vec \cite{mikolov2013efficient}, using large-scale datasets from Mani et al. \cite{mani2018deeptriage} as well as two newly prepared benchmarks: OpenJ9 and TypeScript (TS) (Section \ref{sec:datasets} discusses the datasets). Both TF-IDF and fine-tuned PLMs failed to deliver satisfactory results. \rev{Among prior works, the Knowledge Distillation (KD)-based \cite{hinton2015distillingknowledgeneuralnetwork} model LBT-P showed better performance on our partner's datasets. However, LBT-P can be constrained by the teacher model's generalization limits, affecting performance on small datasets like OpenJ9 \cite{stanton2021does}. Different PLMs tend to capture distinct characteristics, making them better suited to specific types of bug reports \cite{dipongkar2023comparison}. Moreover, PLMs are pretrained on general-domain corpora, and they often rely on shallow cues such as frequent phrasing, leading to reduced robustness on domain-specific tokens (e.g, software-related keywords) \cite{mccoy2019right, elangovan2023effects}. Ensemble methods help mitigate these limitations by combining complementary model strengths \cite{huang2024ensemble, kumar2024ensemble}.}
%(see Section \ref{sec:motivation}).

Based on these observations, we introduce a \textbf{C}ontent-\textbf{B}ased \textbf{R}anker (CBR) in TriagerX, a ranking model for bug triaging that combines embeddings from two PLMs and extracts features from multiple layers. By ensembling these embeddings, CBR leverages complementary information from both models, allowing it to focus on the orthogonality and diversity in the embeddings \cite{bian2022diversity, kuncheva2003diversity}. The final recommendation is made by passing these combined representations through CNN-based classifiers. This approach improved accuracy while being smaller in size than large PLM-based models (see Section \ref{sec:model-efficiency}). CBR outperformed baselines by up to 22\% on literature datasets and 29\% on partner datasets on Top-1 accuracy. %It also achieved 78.20\% Top-1 accuracy across 9 components in the OpenJ9 repository. 
%(see Section \ref{sec:adaptability}).

We also observed that human triagers in our partner teams prioritize developers who recently contributed to similar closed/fixed bugs through commits and comments. Frequent contributors are more likely to be assigned related bugs, yet recent PLM-based methods overlook this. While Yang et al. previously proposed MDN \cite{yang2014recommendation} that exclusively used commit and comment counts for bug triaging, it performed poorly in our partner repositories due to its reliance on the smoothed Unigram Model (UM), which lacks contextual understanding, and its failure to prioritize recent contributions. To address this, we developed the \textbf{I}nteraction-\textbf{B}ased \textbf{R}anker (IBR), which enhances retrieval using a context-sensitive PLM and prioritizes developer contributions in a time-sensitive manner.

\begin{custombox}[colback=gray!10, label=listing:openj9-example, colframe=gray,
label={listing:example-bug-openj9}, title=\small{Listing 1. An example bug report from OpenJ9}
\href{https://github.com/eclipse-openj9/openj9/issues/19231}{(\#19231)}]
\small\textbf{Bug Title:} Build-test doesn't support custom aqa-test repo.\\
\small\textbf{Description:} 
Launching a build via the Jenkins pipeline doesn't allow passing in a custom
aqa-test repo. This complicates testing against excluded tests, requiring
additional effort to ensure comprehensive test coverage without introducing
regressions. FYI \texttt{@llxia} \\
%\small\textbf{Owner:} llxia
\end{custombox}  

In the OpenJ9 repository, IBR outperformed CBR by 4.5\% (Top-1) and exceeded the best literature baseline by 34\%. In TypeScript, IBR ranked second after CBR. This is due to repository dynamics—OpenJ9 has a stable maintainer group with frequent interactions, making past contributions more predictive. In contrast, TypeScript has a diverse, sporadic contributor base, reducing the impact of historical interactions. 

\begin{figure}[t]
\centering
\includegraphics[width=0.85\linewidth]{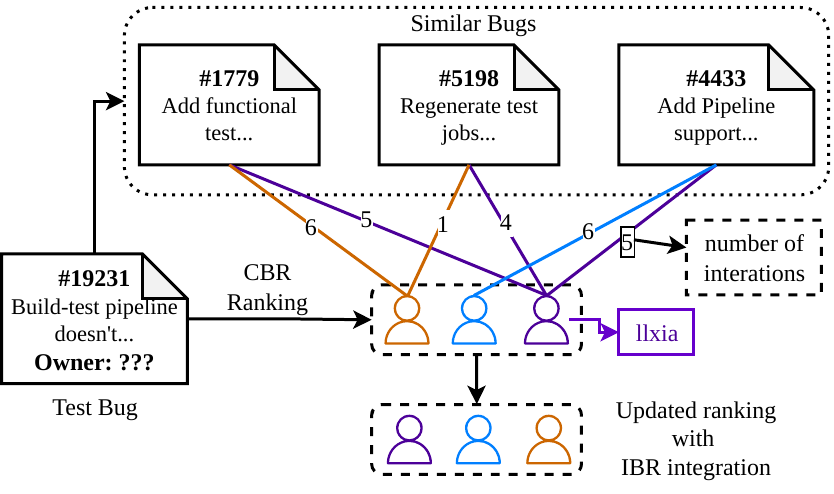}
    \caption{TriagerX updated ranking with IBR for Listing 
    \ref{listing:openj9-example}.}
    \label{fig:triagerx-ibr-example}
    % \vspace{-1em}
\end{figure}

Finally, we combined both CBR and IBR in \textbf{TriagerX} as a hybrid bug triaging framework that ranks developers using CBR and refines the ranking with IBR. To our knowledge, this is the first method to integrate interaction-based ranking with a PLM-driven approach. This version of TriagerX outperforms all methods, achieving Top-1 accuracy improvements of 54\% and 26\% over the best literature baselines, and 58\% and 10\% over the nearest large PLM models for OpenJ9 and TypeScript. For example, in Listing \ref{listing:openj9-example}, despite \texttt{@llxia} being mentioned as the fixer, both LBT-P and standalone CBR in TriagerX misclassified the developer. However, TriagerX, combining CBR and IBR, correctly identified \texttt{@llxia} by considering interaction scores led to the correct prediction, as shown in Figure \ref{fig:triagerx-ibr-example}. 

\revm{In summary, this paper makes the following contributions:}

\begin{itemize}
    \item \revm{\textbf{Dual-transformer content-based ranking.} We propose a novel Content-Based Ranker (CBR) that ensembles multiple pretrained language models and exploits multi-layer representations to capture complementary semantic signals from bug reports for triaging.}
    
    \item \revm{\textbf{Interaction-aware ranking and hybrid integration.} We design an Interaction-Based Ranker (IBR) that models developers’ historical activities through similarity-aware and time-decayed scoring, and integrate it with CBR via a unified hybrid ranking framework that balances textual relevance and developer interaction signals.}
    
    \item \revm{\textbf{Real-world industrial deployment and validation.} We deploy TriagerX within IBM’s OpenJ9 development workflow, demonstrating practical feasibility with low latency and strong real-world acceptance rates.}
\end{itemize}

\section{TriagerX: The Bug Triaging Framework}
\label{sec:methodology}
\begin{figure}[t]
\centering
\includegraphics[width=0.85\linewidth]{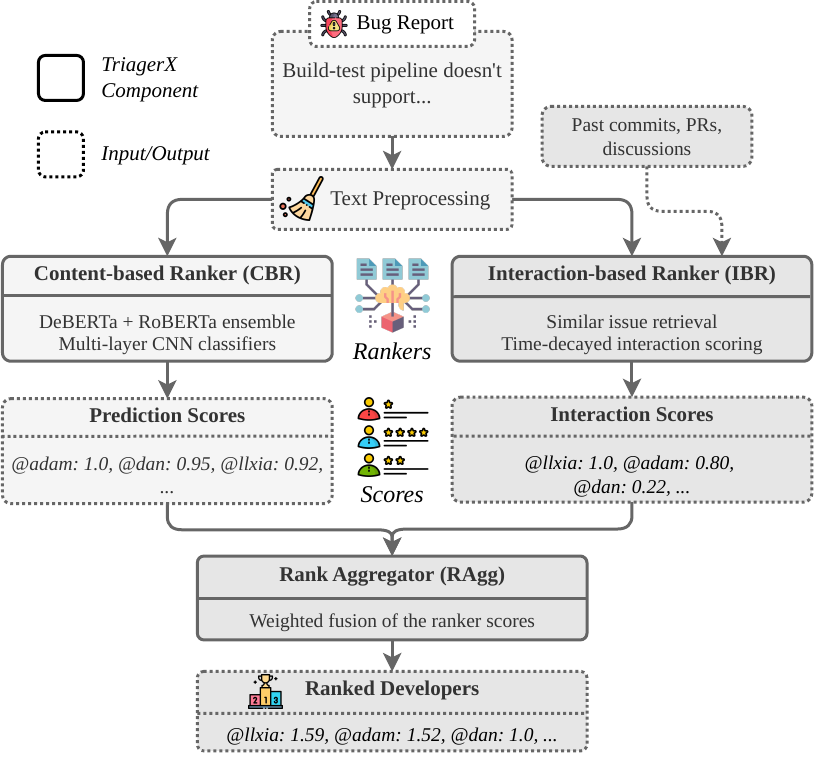}
    \caption{Components and data flow of the TriagerX framework. A bug report (OpenJ9 \#19231) is processed by CBR and IBR in parallel, producing prediction scores and interaction scores, respectively, which are fused by the Rank Aggregator (RAgg) to produce the final ranked developer list.}
    \label{fig:trx-component}
    % \vspace{-1.5em}
\end{figure}

TriagerX has three major components: the Content-based Ranker (CBR), the Interaction-based Ranker (IBR), and the Rank Aggregator (RAgg). The CBR utilizes an ensemble of smaller PLMs to match the performance of larger models while reducing parameters. The IBR employs a scoring method based on developers' historical interactions (e.g., commits, pull requests, discussions/comments, etc.). Finally, the RAgg component applies an effective rank aggregation method to combine both CBR and IBR ranking scores, improving overall recommendation accuracy. \revm{Figure~\ref{fig:trx-component} illustrates the components and data flow of TriagerX using the bug report from Listing~\ref{listing:example-bug-openj9}, showing how the Prediction Scores from CBR and the Interaction Scores from IBR are fused by the RAgg component to produce the final ranked list of developers.}

% \vspace{-1em}

\subsection{Content-based Ranker}
\subsubsection{Overall Architecture}
\begin{figure}[t]
\centering
\includegraphics[width=\linewidth]{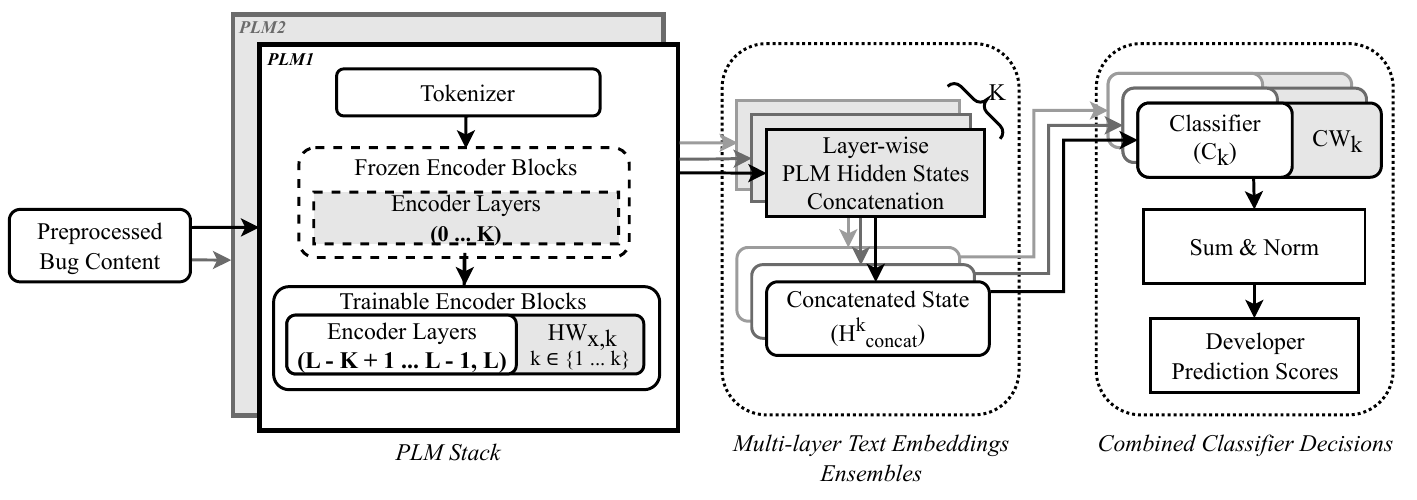}
    \caption{CBR with 2 embedding modules and 3 classifiers.}
    \label{fig:main-arch}
    \vspace{-1em}
\end{figure}

The Content-based Ranker illustrated in Figure \ref{fig:main-arch} consists of two parts: PLMs and classifiers. Multiple PLMs extract rich semantic information from the textual representations of the bug data, providing diverse perspectives on the same bugs as discussed in Section \ref{sec:intro}. \revth{Specifically, we select the base variants of DeBERTa and RoBERTa for CBR due to their overall performance compared to the combination of other PLMs (Table \ref{table:plm-comparison}).}

We concatenate outputs from the last few hidden layers of 2 PLMs and pass them to classifiers, leveraging the hierarchical nature of PLM representations to capture various syntactic and semantic features. While the \texttt{[CLS]} token (a summarized/pooled representation of the input) from the final layer \( H^{(L)} \) provides a context-rich representation, using multiple layers \( H^{(L-K)}, \dots, H^{(L)} \) offers a broader feature set and is found to be more effective for downstream tasks \cite{devlin2018bert, lee2023lbtp}.
To account for varying contributions of hidden states, we use learnable weights \( HW_{x, k} \) for each state, allowing adaptive weighting during classification. The weighted states from all PLMs are concatenated using Equation \ref{eq:rep-concat}.
\begin{equation}
\label{eq:rep-concat}
H_{concat} = \left( \big\|_{i=1}^{x} \text{PLM}_i(L-j) \times HW_{i, j} \right)_{j=0}^{K-1}
\end{equation}

where \( L \) is the total number of layers, \( K \) is the number of classifiers, and \( x \) is the number of PLMs. Section \ref{sec:rq-multiple-plm} discusses how using multiple PLMs enhances TriagerX CBR accuracy.
Each classifier \( C_k \) processes its respective concatenated representation \( H^{(k)}_{concat} \) to produce a prediction \( \hat{y}_k \), which is weighted by learnable parameter \( CW_{k} \) that allows the model to adaptively learn and optimize classifier contributions during training for improved prediction. The combined prediction for CBR is calculated by Equation \ref{eq:cbr-calculation}.
\begin{equation}
\label{eq:cbr-calculation}
{\hat{y}}_{CBR} = \sum_{k=1}^{K} CW_{k} \times F(H^{(k)}_{concat})
\end{equation}
where \( K \) is the number of classifiers and \revm{$F(\cdot)$ is the CNN-based classifier function described in Section \ref{sec:classifier}}. This soft voting approach combines classifier outputs to make the final prediction, integrating diverse representations and improving model robustness \cite{athar2021sentiment, kumari2021ensemble}.

\subsubsection{PLM Fine-Tuning}
Fully fine-tuning a PLM can cause catastrophic forgetting \cite{french1999catastrophic}, while freezing the PLM may lead to underfitting. Early stopping may prevent overfitting but limits learning potential. To address these challenges, we adopt Knowledge Preservation Fine-Tuning (KPFT) \cite{lee2023lbtp}, which selectively freezes the initial layers to retain general language understanding and adapts the later layers to task-specific data. This balances the retention of prior knowledge with the flexibility to learn new representations.
In our setup, for a PLM with \( L \) total layers, during KPFT we define the training status of each layer \( \theta_l \) as:
\begin{equation}
    \theta_l = \begin{cases} 
  \text{trainable}, & \text{if } l \geq L - K + 1 \\
  \text{frozen}, & \text{if } l < L - K + 1 
\end{cases}
\end{equation}
where \( K \) denotes the number of classifiers being trained. By freezing the majority of the PLM's encoder layers and selectively updating only those necessary for the downstream task, KPFT ensures that each classifier can effectively utilize task-specific representations while preserving the broader linguistic knowledge encoded in the initial layers.

% \begin{wrapfigure}[18]{r}{0.35\linewidth}
%     \centering
%     \vspace{-1.2em}
%     \includegraphics[width=0.9\linewidth]{figures/classifier.pdf}
%     \caption{Classifier architecture of TriagerX CBR.}
%     \label{fig:classifier}
% \end{wrapfigure}
\subsubsection{Developer Classifier}
\label{sec:classifier}
\begin{figure}[t]
\centering
\includegraphics[width=\linewidth]{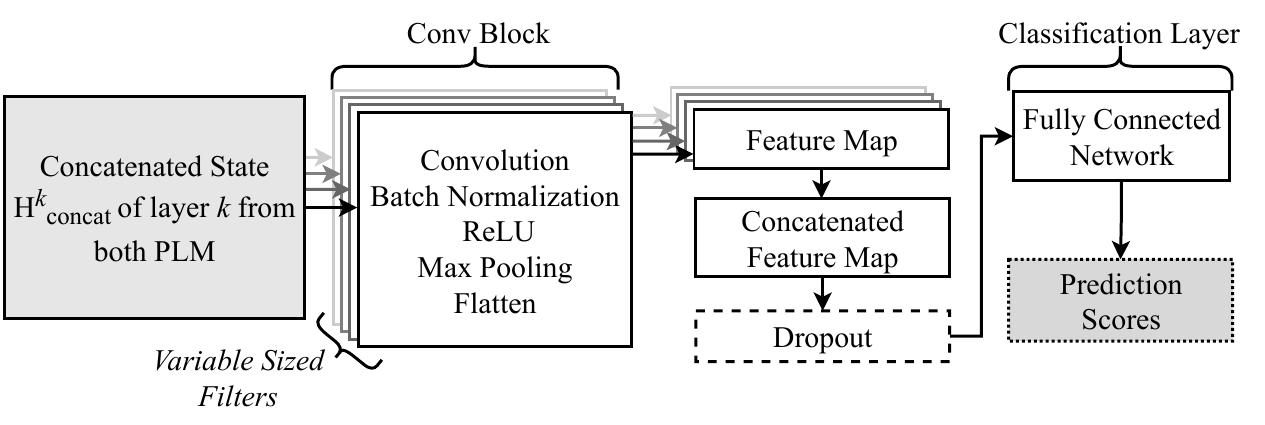}
    \caption{Classifier architecture of TriagerX CBR.}
    \label{fig:classifier}
    \vspace{-1em}
\end{figure}

Fully connected networks (FCNs) are generally used on PLM embeddings. However, our experiments show that simple FCNs may not provide optimal results for bug-triaging task (Section \ref{sec:rq4-cbr-design}). To improve classification accuracy, we adopt a CNN-based architecture, visualized in Figure \ref{fig:classifier}. CNNs capture local patterns and hierarchical features more efficiently, making them suitable for processing concatenated hidden states from the PLMs. \revtwo{Following prior work, each convolution block (CB) uses four parallel convolution layers with filter sizes \( f_s \in \{3, 4, 5, 6\} \), each with 256 filters allowing the model to capture features of different n-gram lengths \cite{lee2023lbtp}.} Each convolution layer is followed by batch normalization (BN) \cite{ioffe2015bn}, ReLU activation, max pooling and flattening.
Finally, we concatenate the feature maps from the parallel convolution layers. Before passing the feature vector to the FCN for the recommendation task, we apply a dropout layer \cite{srivastava14a2014dropout} to minimize overfitting by randomly setting a fraction of inputs to zero during training.
% \vspace{-1em}    

\subsection{Interaction-based Ranker}
% The proposed CBR significantly outperforms existing methods using only bug titles and descriptions. 
% As discussed in Section \ref{sec:intro}, even well-trained models can perform poorly. To improve accuracy, TriagerX integrates developer interactions such as assignments, commits, pull requests, discussions, and timestamps. This heuristic-based approach accounts for recent activities, enhancing predictions by incorporating temporal information.
We introduce a novel interaction scoring technique
%algorithm (Algorithm \ref{alg:contribution-scoring}) 
to evaluate developers based on past involvement with similar bugs. Each developer receives an interaction score (IS) from their historical contributions, which adjusts the prediction scores from CBR to improve accuracy. 
%\subsubsection{IBR Ranking Algorithm}
The IBR technique begins by identifying similar issues using Siamese BERT networks (SBERT) \cite{reimers-2019-sentence-bert}. We leverage a pre-trained SBERT model from the official repository as these models provide strong text similarity capabilities off-the-shelf without requiring additional fine-tuning. SBERT efficiently encodes sentences into dense vector representations, allowing fast and accurate retrieval of similar issues via cosine similarity, as defined in Equation \ref{eq:cos-sim}.
\begin{equation}
    \label{eq:cos-sim}
    \text{Similarity}(i, j) = \frac{\mathbf{e}_i \cdot \mathbf{e}_j}{\| \mathbf{e}_i \| \| \mathbf{e}_j \|}
\end{equation}
\revm{where $\mathbf{e}_i$ and $\mathbf{e}_j$ are the dense vector embeddings of issues i and j, respectively, and issues i and j are considered similar if their similarity meets the threshold $\tau$:}

\begin{equation}
% \hspace{-16em}
    \text{IsSimilar}(i, j) = 
    \begin{cases} 
    \text{similar,} & \text{if } \text{Similarity}(i, j) \geq \tau \\
    \text{dissimilar,} & \text{if } \text{Similarity}(i, j) < \tau 
    \end{cases}
\end{equation}

Since SBERT models generate context-sensitive embeddings independent of term frequencies between reports unlike \cite{yang2014recommendation}, embeddings for all existing issues can be precomputed and stored, enabling real-time and accurate retrieval of similar issues. Once retrieved, timeline events of each similar issue are traversed to identify developers who interacted with it through commits, pull requests, discussions, or assignments. Each interaction is scored differently based on its type. For example, a commit to an issue may signify a developer's involvement more than a comment. Contributions from \textit{inactive developers} are excluded by maintaining a directory of \textit{active developers (D)}. We define an active developer as an individual who contributes a minimum number of times to a repository. 

\revm{To account for temporal relevance, we weigh developer interactions using an exponential decay function \( f(t) = e^{-\lambda t} \), where \( \lambda \) is the decay rate and \( t \) is the interval between the interaction date and the report time of the new issue being triaged, so that only interactions from issues reported before the new issue are considered.} This approach ensures that recent contributions have a stronger influence on the ranking, while older interactions are smoothly and progressively discounted. Intuitively, a developer who interacted with a similar issue just two days ago is more likely to work on a related issue than someone who last interacted a year ago, even if both were involved in similar components. \rev{We adopt exponential decay over linear or stepwise alternatives because it offers a smooth, continuous discounting that aligns with how developer engagement typically fades over time. Prior work in collaborative filtering also showed that exponential decay more effectively models evolving user preferences and time-aware relevance \cite{ding2005time, jain2023performance}.} The total interaction score (IS) of a developer \( d_i \) for a new issue ($I_{\text{new}}$) is calculated by Equation \ref{eq:ranking-eq}.
\begin{equation}
    IS(d_i) = \sum_{j=1}^{m} \sum_{k=1}^{n_{ij}} \text{Sim}(I_{\text{new}}, I_{j}) \cdot IP_{ijk} \cdot e^{-\lambda t_{ijk}}
    \label{eq:ranking-eq}
\end{equation}

\revm{where $m$ is the number of similar issues retrieved for the new issue ($I_{\text{new}}$), $n_{ij}$ is the total number of interactions by developer $d_i$ on issue $I_j$, while $\text{Sim}(I_{\text{new}}, I_j)$ is the cosine similarity between $I_{\text{new}}$ and retrieved issue $I_j$ (Equation \ref{eq:cos-sim}) weighting each interaction by the relevance of the issue it occurred on. $IP_{ijk}$ is the interaction point value for the k-th interaction of developer $d_i$ on issue $I_j$, drawn from a predefined table based on interaction type (e.g., commit/PR, discussion, assignment — see Table \ref{table:contribution-scoring-parameters}), $\lambda$ is the time decay rate controlling how quickly older interactions are discounted, and $t_ijk$ is the elapsed time in days between the k-th interaction on issue $I_j$ and the report time of the new issue $I_\text{new}$.} Once scoring is done for all developers, scores are normalized by Equation \ref{eq:normalization}.

% Here, \( IS(d_i) \) is the interaction score for developer \( d_i \in D \), computed as the sum of interaction points (IP) across multiple issues they engaged with. \( IP_{ijk} \) is retrieved from a predefined interaction point table based on the interaction type. \( k \) ranges from 1 to \( n_{ij} \) (total contributions to issue \( I_j \)), weighted by the similarity score \( \text{Sim}(I_{\text{new}}, I_{j}) \) between the new issue \( I_{\text{new}} \) and \revm{existing issue} \( I_j \), reflecting the relevance of the developer’s input. Interaction points are adjusted for recency using \( e^{-\lambda t_{ijk}} \), where \( t_{ijk} \) denotes the days since the \( k \)-th interaction with issue \( I_j \) by developer \( d_i \). Once scoring is done for all developers, scores are normalized by Equation \ref{eq:normalization}.

\begin{equation}
\label{eq:normalization}
    NIS(d_i) = \frac{IS(d_i) - \min(IS)}{\max(IS) - \min(IS)}
\end{equation}

 % We adopt exponential decay over linear or stepwise alternatives due to its continuous nature, lack of abrupt cutoffs, and strong alignment with how developer involvement typically diminishes over time. Prior work in collaborative filtering has shown that exponential decay more accurately captures evolving user preferences than uniform or fixed-window schemes \cite{ding2005time}, and offers superior performance in time-aware similarity computations \cite{jain2023performance}.

% To account for time, interactions from different timestamps are weighted differently using an exponential decay function \( e^{-\lambda t} \), where \( \lambda \) is the decay factor and \( t \) is the time interval between the interaction date and the present date. This ensures recent interactions exert more influence on the developer ranking. We chose exponential decay over linear or stepwise alternatives because it provides smooth, continuous discounting, avoids abrupt cutoffs, and has been shown to better model evolving user behavior. In collaborative filtering, Ding et al. \cite{ding2005time} demonstrated that exponential decay captures shifting interests more effectively than uniform weighting or fixed windows, while Jain et al. \cite{jain2023performance} empirically validated its superior performance in time-aware similarity computations. 

\begin{algorithm}[t]
    \small
    \SetAlgoLined
    \SetKwInOut{Input}{Input}
    \SetKwInOut{Output}{Output}
    
    \Input{New Issue $I_{\text{new}}$, Similarity threshold $ \tau $, Time decay factor $\lambda$, Existing all issues $I_\text{all}$, Existing issues embedding $\mathbf{E} = \{\mathbf{e}_1, \mathbf{e}_2, \dots, \mathbf{e}_N\}$, Interaction point table $IT$, Active developers $D = \{d_1, d_2, \dots, d_n\}$}
    \Output{Normalized interaction scores (NIS)}
    
    \BlankLine
    Compute new issue embedding $\mathbf{e}_{\text{new}} = \text{SBERT}(I_{\text{new}})$\;
    Calculate cosine similarities with existing issues: $\text{Sim}(I_{\text{new}}, I_j) = \frac{\mathbf{e}_{\text{new}} \cdot \mathbf{e}_j}{\|\mathbf{e}_{\text{new}}\| \|\mathbf{e}_j\|}, \forall I_j \in I_{\text{all}}$;
    
    Select top similar issues $I_{\text{top}} = \{ I_j \mid \text{Sim}(I_{\text{new}}, I_j) \geq \tau \}, \forall I_j \in I_{\text{all}}$;
    
    % $I_\text{top}$ =  $\text{Sim}(I_{\text{new}}, I_j) \geq \tau $\;
    
    % \ForEach{issue $I_j$}{
    %     Extract all developers $D_j$ who interacted with $I_j$\;
    % }
    
    Initialize a score dictionary $IS(d_i) = 0, \forall d_i \in D$\;
    
    \ForEach{issue $I_j \in I_\text{top}$}{
        \ForEach{interaction $k$ by developer $d_i$ on issue $I_j$}{
            \If{$d_i \in D$}{
                Get interaction point $IP_{ijk}$ from $IT$\;
                Update $IS(d_i) = IS(d_i) + \text{Sim}(I_{\text{new}}, I_j) \cdot IP_{ijk} \cdot e^{-\lambda t_{ijk}}$\;
            }
        }
    }
    
    \If{\( \exists d_i \in D \text{ such that } IS(d_i) > 0 \)}{
        Normalize Interaction Scores $(NIS)$ using Equation \ref{eq:normalization}

        \Return $NIS$;
    }
    \Return $IS$;
    
    \caption{\revth{IBR developer scoring algorithm based on previous interactions.}}
    \label{alg:contribution-scoring}
\end{algorithm}
% \vspace{-1em}

Normalization is only done if there is at least one developer with a non-zero interaction score to avoid division by zero. The overall algorithm is provided in Algorithm \ref{alg:contribution-scoring}.

\begin{figure}[t]
\centering
\includegraphics[width=0.8\linewidth]{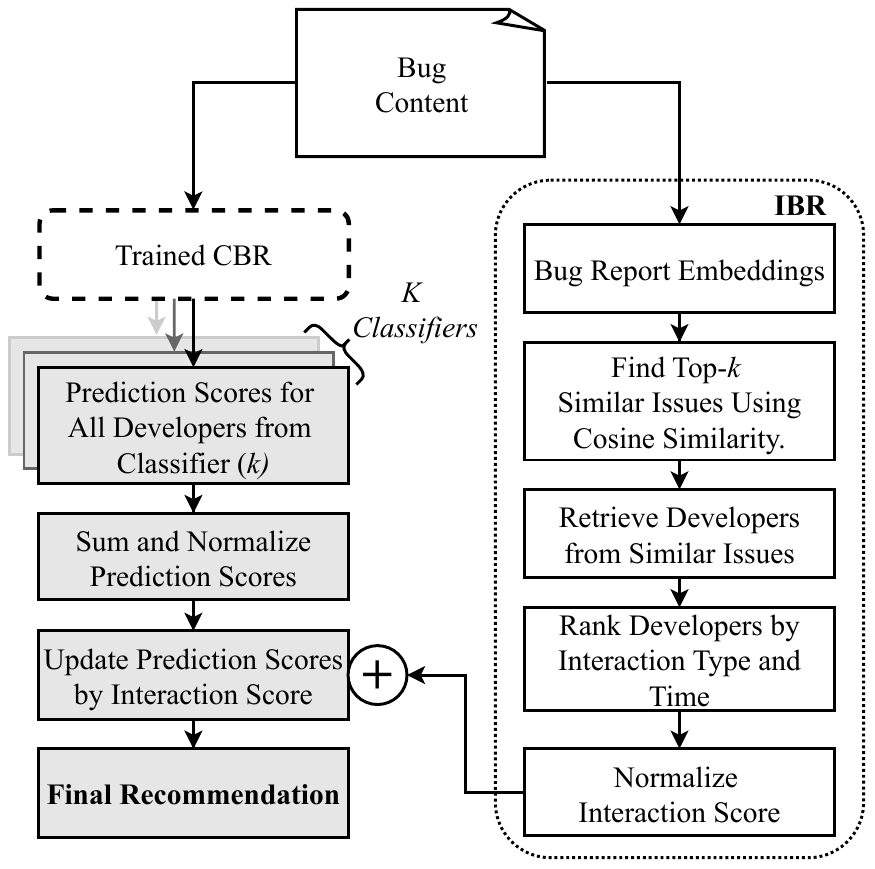}
\caption{TriagerX complete framework with both CBR and IBR.}
\label{fig:triagerx-enhanced}
% \vspace{-1em}
\end{figure}

% The normalized score adjusts the original scores of CBR.

% \begin{figure}[htbp]
%     \centering
%     \includegraphics[width=0.7\linewidth]{figures/triagerx-full-arch.pdf}
%     \caption{TriagerX complete framework with both CBR and IBR.}
%     \label{fig:triagerx-enhanced}
% \end{figure}
% \vspace{-1em}

\subsection{Rank Aggregator}
\label{sec:ragg}
We employ a Weighted Ranking Aggregation (WRA) method that combines the normalized prediction score (NPS) with the weighted normalized interaction score (NIS). Developers who have not interacted to any similar issues retain their original prediction score from TriagerX CBR. The interaction scores are multiplied by a weighting factor \( W_f \), which is optimized through grid search within the range \((0, 1)\) to ensure appropriate weighting of the interaction score depending on the dataset/repository. The calculation of the final score (FS) for developer \( d_i \) is expressed as:
\begin{equation}
\label{eq:rank-agg}
    FS(d_i) = NPS(d_i) + W_f \cdot NIS(d_i)
\end{equation}

Figure \ref{fig:triagerx-enhanced} shows the complete architecture of the TriagerX framework, including both CBR and IBR.

\section{Evaluation}\label{sec:evaluation}
\noindent Our goal is to study whether and how TriagerX outperforms the SOTA tools. We answer five research questions (RQs):
\begin{enumerate}[label=\textbf{RQ}\textbf{\arabic{*}.}, leftmargin=28pt]
    \item Can TriagerX outperform the baselines?
    \item How effective are the ranker components individually?
    \item How do multiple PLMs influence the accuracy of CBR?
    \item Do the design decisions impact CBR performance?
    \item Do hyperparameter choices affect IBR performance?
\end{enumerate}
RQ1 and RQ2 evaluate the framework's accuracy and check the statistical significance of the results, with RQ1 focusing on overall accuracy and RQ2 assessing individual ranker components. RQ3 examines the effectiveness of combining multiple PLMs. RQ4 and RQ5 explore different design choices in the overall framework in the form of an ablation study.

\vspace{-1em}
\subsection{Experimental Setup}
% \subsubsection{PLM Selection}
% We evaluated high-performing PLM variants for bug triaging, as reported in \cite{dipongkar2023comparison}. All PLMs were sourced from the HuggingFace repository \cite{huggingface} and served as the core text embedding modules. We used the large variants of BERT \cite{devlin2018bert}, RoBERTa \cite{liu2019roberta}, and DeBERTa \cite{he2024deberta}, given their typical superiority over base versions, along with the base variant of CodeBERT \cite{feng-etal-2020-codebert}, as it is the only publicly available option. For TriagerX CBR, we fine-tuned the base versions of RoBERTa and DeBERTa in an ensemble, showing that with careful tuning and model combination, base models can outperform larger ones in bug triaging.

\subsubsection{Benchmark Datasets}
\label{sec:datasets}

\revth{We utilized large-scale Google Chromium (GC), Mozilla Core (MC), and Mozilla Firefox (MF) datasets from Mani et al. \cite{mani2018deeptriage}. These datasets include issue title, description, reporting date, and the owner who resolved the bug. As the literature datasets do not contain developer interaction information, we also prepared two new benchmark datasets for our analysis. To create our own benchmarks, we leveraged the GitHub API to collect data from the OpenJ9 and TypeScript (TS) bug repositories. We gathered all reported bugs up to June 2024, dating back to the creation of each repository. These datasets include information such as issue titles, descriptions, assigned developers, and contributors (e.g., those who commented, committed, or created pull requests). Directly assigned developer to an issue is considered the owner of a bug report. In cases where there is no direct assignment on the GitHub issue page, we considered the last person to make a commit or pull request to that bug as the owner. If neither of this information was found on the issue page, we discarded the issue from our dataset.}

\noindent\textbf{Text Preprocessing.} We applied consistent preprocessing across all datasets. First, we concatenated the issue title and description, then filtered out reports with fewer than 15 words to ensure sufficient context. URLs, timestamps, special characters, and code block identifiers (e.g., \texttt{```}) were removed. Hexadecimal values were replaced with \texttt{<hex>} tokens to maintain log structure without introducing noise. For example, "Memory error at 0x7fff5fbff7a0" was transformed to "Memory error at \texttt{<hex>}." We used dynamic tokenization to ensure each PLM uses its respective pretrained tokenizer. Texts were tokenized to a maximum of 256 tokens, and shorter texts were padded. %using \texttt{<PAD>}.

\noindent\textbf{Sampling Active Developers.} In most software repositories, a small subset of developers contribute frequently, while the majority have limited, sporadic contributions—especially in open-source projects. To improve ML model generalization, often, only frequent contributors, termed \textit{active developers} \cite{anvik2006svm, park2016cost, hadi2023adptriage}, are considered. We define an active developer \( D \) as one with contributions \( C(D) \geq 20 \), excluding those with \( C(D) < 20 \) to reduce outliers from insufficient training data. The threshold of 20 is motivated by Mani et al. \cite{mani2018deeptriage}, who observed that deep learning models show improved accuracy as the threshold increases up to 20, after which gains plateau. This filtering accounts for the difference in the number of issues and active developers shown in Table \ref{table:dataset-summary} after preprocessing, compared to one of our baselines, LBT-P \cite{lee2023lbtp}.

% \begin{wraptable}{c}{2.5in}
% \small
% \vspace{-2.5em}
% \caption{Datasets after preprocessing.}
% \centering
% \begin{tabular}{lccc}
% \toprule
% \textbf{Dataset} & \textbf{Train Bug} & \textbf{Test Bug} & \textbf{\shortstack{Active \\ Developers}}\\
% \midrule
% Google Chromium & 98,165 & 10,781 & 986 \\
% Mozilla Core & 108,607 & 10,994 & 557 \\
% Mozilla Firefox & 18,223 & 1,755 & 155 \\
% Openj9 & 3,375 & 382 & 50 \\
% Typescript & 9,929 & 1,129 & 40 \\
% \bottomrule
% \end{tabular}
% \label{table:dataset-summary}
% \end{wraptable}

% \begin{table}
% \begin{center}
% \caption{Datasets after preprocessing.}
% \label{table:dataset-summary}
% \resizebox{\linewidth}{!}{
% \begin{tabular}{lccccc}
% \toprule
% \textbf{Dataset} & \textbf{\revth{Time Range}} & \textbf{Train Bug} & \textbf{Test Bug} & \textbf{Active Dev} & \textbf{Max Contrib.}\\
% \midrule
% Google Chromium & \revth{08/2008 -- 07/2016} & 98,165 & 10,781 & 1032 &  \\
% Mozilla Core & \revth{04/1998 -- 06/2016} & 108,607 & 10,994 & 557 &  \\
% Mozilla Firefox & \revth{07/1999 -- 06/2016} & 18,223 & 1,755 & 155 & \\
% Openj9 & \revth{09/2017 -- 06/2024} & 3,375 & 382 & 50 & \\
% Typescript & \revth{07/2014 -- 06/2024} & 9,929 & 1,129 & 40 & \\
% \bottomrule
% \end{tabular}
% }
% % \vspace{-2.5em}
% \end{center}
% \end{table}

\begin{table}
\begin{center}
\caption{Datasets after preprocessing. Mean$^\dagger$ and Std$^\dagger$ refer to the average and standard deviation of issues owned per active developer.}
\label{table:dataset-summary}
\resizebox{\linewidth}{!}{
\begin{tabular}{lcccccc}
\toprule
\textbf{Dataset} & \textbf{\revth{Time Range}} & \textbf{Train} & 
\textbf{Test} & \textbf{Devs} & \textbf{Mean$^\dagger$} & \textbf{Std$^\dagger$}\\
\midrule
GC & \revth{08/2008 -- 07/2016} & 98,165  & 10,781 & 1,032 & 110.5 & 115.4 \\
MC    & \revth{04/1998 -- 06/2016} & 108,607 & 10,994 & 557   & 214.7 & 342.3 \\
MF & \revth{07/1999 -- 06/2016} & 18,223  & 1,755  & 155   & 128.9 & 167.6 \\
OpenJ9          & \revth{09/2017 -- 06/2024} & 3,375   & 382    & 50    & 74.0  & 76.4  \\
TS      & \revth{07/2014 -- 06/2024} & 9,929   & 1,129  & 40    & 276.2 & 353.6 \\
\bottomrule
\end{tabular}
}
\end{center}
\end{table}

\noindent\textbf{\revm{Dataset Characteristics.}} \revm{We analyze the characteristics of all datasets after preprocessing, with key statistics summarized in Table~\ref{table:dataset-summary}. Mozilla Core shows the highest variability (mean: 214.7, std: 342.3), driven by a small number of extremely active developers, while OpenJ9 is the most balanced (mean: 74.0, std: 76.4). TypeScript shows a similar degree of skewness to Mozilla Core (mean: 276.2, std: 353.6) despite having only 40 active developers, with the top 5 developers alone accounting for 48.3\% of all resolved issues. We additionally analyze label diversity for our benchmark datasets (TypeScript and OpenJ9) as the literature datasets sourced from Mani et al.~\cite{mani2018deeptriage} do not include label or interaction metadata beyond issue titles, descriptions, and developer assignments. TypeScript issues carry an average of 2.27 labels per issue across 127 unique labels (e.g., \textit{Bug}, \textit{Suggestion}), reflecting a broad and diverse issue taxonomy. OpenJ9 issues carry an average of 2.17 labels across 84 unique labels, dominated more by component-specific labels such as \textit{comp:vm} and \textit{comp:jit}, reflecting a more structured labeling scheme consistent with OpenJ9's stable maintainer group.}

\subsubsection{Content-based Ranker Training}
\label{sec:cbr-training}
We trained the CBR model across all datasets, using three classifier blocks on the concatenated hidden states from the final three encoder layers of the PLMs. Following literature baselines \cite{lee2023lbtp, mani2018deeptriage}, datasets were split chronologically, with 90\% for training and 10\% for testing. Although we focus on active developers, contribution levels vary significantly. For instance, in the GC dataset, the top contributor has over 1100 contributions, while the lowest after filtering has 20, creating class imbalance. To address this, we used weighted random sampling, assigning each issue a weight inversely proportional to its developer’s frequency:

\begin{equation}
    P_i = \frac{1 / f_i}{\sum_{j=1}^{n} (1 / f_j)}
\end{equation}

where \( P_i \) is the probability of selecting issue \( i \), \( f_i \) is the frequency of issues assigned to developer \( d_i \), and \( n \) is the total number of issues. This approach balances the training data by sampling issues associated with less frequent developers more often. \revtwo{We used the AdamW optimizer \cite{ilya2017adamw} with a weight decay of \(0.001\) and a linear learning rate schedule, including a \(1 \times 10^{-5}\) peak learning rate and 10\% warmup.} Models were trained for up to 40 epochs on an NVIDIA A100 GPU.

\subsubsection{Interaction-based Ranker Optimization}
\label{sec:ibr-optimization}
TriagerX IBR uses the pre-trained SBERT model \textit{all-mpnet-base-v2} \cite{song-2020-mpnet} for embedding generation without requiring task-specific fine-tuning, chosen for its superior accuracy over other models. Detailed comparisons between these models are not provided as they are not the focus of our research. For the experimental setup, embeddings for all training issues are pre-computed. When a new bug is reported, similar bugs from the existing issues are identified based on embedding similarity. Interaction scores are subsequently calculated by considering the nature of interactions with similar issues, including both interaction type and timing.

\noindent\textbf{Optimization Process.} To adapt interaction scoring to different repository settings, IBR avoids relying on fixed heuristic values and instead performs a straightforward hyperparameter tuning step to adjust key parameters (e.g., $\lambda$, $IP_x$) for optimal performance. Given the small number of parameters and a constrained search space, we employ an exhaustive grid search over all possible combinations of parameter values. \rev{Optimization is performed using a validation set sampled from less than 10\% of the training data.} Let \( f(\mathbf{P}) \) represent the accuracy of the framework evaluated at a specific parameter set \( \mathbf{P} \) within the defined grid of parameters $\mathcal{P} = \{ (\tau, \lambda, IP_x, W_f) \mid \tau \in [\tau_{\text{min}}, \tau_{\text{max}}], \lambda \in [\lambda_\text{min}, \lambda_\text{max}] \ldots \}$. The optimal parameter set is determined as follows:

\begin{equation}
\label{eq:grid-optim}
\mathbf{P}_{\text{best}} = \arg\max_{\mathbf{P} \in \mathcal{P}} f(\mathbf{P}),
\end{equation}

where \( \mathbf{P}_{\text{best}} \) is the parameter set that maximizes the accuracy \( f(\mathbf{P}) \) across the grid \( \mathcal{P} \). Each dataset may require specific parameter adjustments due to unique characteristics like bug type distributions and developer activity patterns. We discuss these parameters and their search space below:

\noindent\textbf{Similarity Threshold \( \tau \in [0.2, 0.8] \)}. \rev{Filters semantically similar issues based on cosine similarity. Scores below 0.2 are typically unrelated and thus excluded. Conversely, issues with similarity above 0.8 are highly likely to be relevant. Hence, we limit the upper bound of the search range to 0.8 to reduce the search space.} \revm{To ensure computational efficiency, similar issue retrieval is capped at a maximum of 20 issues per query. This prevents unbounded costs in both retrieval and subsequent interaction score calculation from bug reports at low threshold values explored during grid search or in deployment to repositories with large numbers of existing issues.}

% \noindent\textbf{Maximum Similar Issues \( K \in [5, 20] \)}: Controls how many semantically similar past issues are considered when computing interaction scores. While the similarity threshold \( \tau \) filters out irrelevant examples, \( K \) bounds the total number of candidates, preventing excessive aggregation and reducing noise from marginally similar bugs. We set the range from 5 to 20 to ensure sufficient historical context while maintaining computational efficiency. 

\noindent\textbf{Time Decay Factor \( \lambda \in [0.001, 0.01] \)}. \rev{This parameter controls how quickly interaction scores decay over time. Similar to prior research \cite{ding2005time}, we allow this wide range to adapt to different repository dynamics flexibly.}

\noindent\textbf{Interaction Points \( IP_x \in [0, 2] \)}. \rev{Allows the framework to learn the relative importance of different developer interactions—assignments, commits/PRs, and discussions. The range is designed to represent three interpretable levels (0: low, 1: medium, 2: high), with fractional values (e.g., 0.5, 1.5) enabling finer control. As shown in Table~\ref{table:contribution-scoring-parameters}, grid search typically assigns higher weights to commits/PRs than to discussions, reflecting their greater predictive value in developer activity.}

\noindent\textbf{Interaction Score Weight \( W_f \in (0, 1) \)}. \rev{\( W_f \) controls the contribution of IBR relative to CBR during final rank aggregation (Equation \ref{eq:rank-agg}). This parameter allows the framework to adjust the weight of IBR ranking scores. For example, a lower weight is found to be more effective in repositories with sparse interactions (e.g., TS), while higher weights benefit repositories with dense interaction histories (e.g., OpenJ9).}

\rev{Table~\ref{table:contribution-scoring-parameters} presents the optimized hyperparameters for the OpenJ9 and TS datasets, along with their corresponding search ranges and step size (i.e., interval between consecutive values explored) used during the tuning process.}

% \vspace{-1em}
% \begin{wraptable}{r}{7cm}
\begin{table}
\small
% \vspace{-2em}
\caption{Optimized hyperparameters for interaction-based ranker on different datasets obtained via grid search.}
\centering
\resizebox{\linewidth}{!}{
\begin{tabular}{lcccc}
\toprule
\textbf{Parameter} & \textbf{Openj9} & \textbf{TS} & \textbf{Search Space} & \textbf{\rev{Step}}\\
\midrule
Similarity Threshold $ \tau $ & 0.4 & 0.45 & [0.2, 0.8] & \rev{0.05}\\
% Maximum Similar Issues $K$ & 20 & 20 & [5, 20]\\
Time Decay Factor $\lambda$ & 0.01 & 0.001 & [0.001, 0.01] & \rev{0.001} \\
Interaction Point (Assignee) $IP_a$ & 0.5 & 0.5 & [0, 2] & \rev{0.10} \\
Interaction Point (Commit/PR) $IP_c$ & 1.5 & 1.5 & [0, 2] & \rev{0.10} \\
Interaction Point (Discussion) $IP_d$ & 0.20 & 0.10 & [0, 2] & \rev{0.10} \\
% \midrule
Interaction Score Weight $W_f$ & 0.65 & 0.25 & (0, 1) & \rev{0.05} \\
\bottomrule
\end{tabular}}
\vspace{-1em}
\label{table:contribution-scoring-parameters}
\end{table}

\subsection{SOTA Baselines} 

% We benchmarked TriagerX against the baselines introduced in Section \ref{sec:intro} to provide a comprehensive evaluation. LBT-P \cite{lee2023lbtp} was chosen as the latest PLM-based SOTA model, DBRNN-A \cite{mani2018deeptriage} for its use of context-free embeddings, and MDN \cite{yang2014recommendation} for its focus on developer interactions, such as commits and comments. Additionally, we included traditional TF-IDF-based methods and larger PLM architectures with both FCN and CNN classifiers for a well-rounded comparison. See Appendix \ref{sec:app-baselines} for more details on baselines.

\revth{We used the following approaches as our baselines.}

% \noindent\textbf{LBT-P.} \revth{is a bug triage framework using patient knowledge distillation (PKD) approach to compress RoBERTa-large that attempts to mitigate PLM's overthinking problem \cite{kaya2019shallow}. As LBT-P’s source code was not publicly available, we contacted the authors, who informed us that it was proprietary and therefore could not be shared. Hence, we reproduced the framework by meticulously following the paper's methodologies. Initial discrepancies in results prompted us to communicate with the authors again, who then provided some partial code snippets and recommended a higher learning rate of \(1e^{-4}\) for distillation. With these adjustments, we reproduced similar results to the original on the GC dataset, validating our implementation. However, reproducing results on MC and MF was challenging due to our use of \textit{active developers} and significant data distribution shifts. To better evaluate model performance, we maximized developer overlap between training and test sets, reducing data sparsity for a more realistic assessment of generalization. This adjustment impacted all baselines and our approach consistently, ensuring fair comparisons.}

\noindent\textbf{LBT-P.} \revth{is a bug triage framework using patient knowledge distillation (PKD) approach to compress RoBERTa-large that attempts to mitigate PLM's overthinking problem \cite{kaya2019shallow}. As the official source code was not publicly available, we contacted the authors and were informed that the implementation is proprietary. Nevertheless, they shared partial code snippets and detailed reproduction guidelines. Following the paper and the authors’ instructions, we reimplemented the framework. On the GC dataset, our reproduction achieved results comparable to those reported in the original study, providing confidence in the correctness of our implementation.}

\noindent\textbf{DBRNN-A.} \revth{is a deep bidirectional RNN with Attention that uses LSTM units to capture context in bug reports, addressing the challenge of mixed text, code snippets, and stack traces. Our reproduction of DBRNN-A following this \href{https://github.com/hacetin/deep-triage}{implementation} yields similar results to the original paper. }

\noindent\textbf{MDN.} \revth{Multiple Developer Network first attempts to find similar issues by applying smoothed Unigram Model (UM) and Kullback-Leibler (KL) divergence. Then it generates a network of developers by the number of commits and comments on those bug reports.}

\noindent\textbf{Large PLMs with FCN and CNN classifiers.} \revth{We evaluated high-performing PLM variants for bug triaging, as reported in \cite{dipongkar2023comparison}. All PLMs were sourced from the HuggingFace repository \cite{huggingface} and served as the core text embedding modules. We used the large variants of BERT \cite{devlin2018bert}, RoBERTa \cite{liu2019roberta}, and DeBERTa \cite{he2024deberta}, given their typical superiority over base versions, along with the base variant of CodeBERT \cite{feng-etal-2020-codebert}, as it is the only publicly available option.
For FCN-based classifiers, we used the final-layer \texttt{[CLS]} token representation as the sequence-level embedding. For CNN-based classifiers, we applied convolutional filters (same CNN configuration as CBR without ensembling) over the hidden states from the last transformer layer, followed by pooling to obtain a fixed-size representation.}   

\noindent\textbf{TF-IDF+SVM.} \revth{We use term frequency–inverse document frequency (TF-IDF) vectors to represent bug report text, capturing word-level importance across the corpus. These features are then fed into a linear Support Vector Machine (SVM) classifier, serving as a non-neural baseline for bug triaging.}

\revth{We also evaluated recent graph-based bug triaging methods, including NCGBT \cite{dong2024neighborhood} and PCG \cite{dai2024pcg}, using their official implementations.} \revm{Both approaches performed poorly on our datasets. NCGBT relies on CC fields that are absent in our data, which likely explains its sharp performance drop as the dataset size increases, achieving only 0.035 Top-1 accuracy on OpenJ9 and $\approx$0.004 on Google Chromium. Similarly, while PCG is not dependent on CC data, it exhibited very low Top-k accuracy on large projects, consistent with the original paper's report for Google Chromium (Top-1: 0.008) \cite{dai2024pcg}. For completeness and transparency, we include these results in Appendix \ref{app:additional-baselines} and provide reproduction artifacts and hyperparameter configurations for these methods in our replication package, but omit them from the main paper due to their inferior performance on our datasets compared to many other baselines.}

% \vspace{-1.5em}
\subsection{Evaluation Metric}
% Following our literature baselines \cite{mani2018deeptriage, lee2023lbtp}, we define bug triaging as a classification problem where each incoming bug report includes a title, a descriptive body (which may contain text, stack traces, code snippets, error logs, and reproduction steps). The goal is to assign the bug to the most suitable developer. 
Following our literature baselines \cite{mani2018deeptriage, lee2023lbtp}, we employ the Top-k metric to evaluate performance. Multiple developers may be qualified to fix a bug. Therefore, Top-k accuracy is defined as:

\begin{equation}
\text{Top-k Accuracy} = \frac{1}{N} \sum_{i=1}^{N} \mathbb{I} \left( y_i \in F(x_i, k) \right)
\end{equation}

\( N \) is the number of bugs, \( y_i \) is the actual developer assigned to bug \( i \), \( x_i \) is the bug's input, \( F(x_i, k) \) is the set of Top-k predicted developers, and \( \mathbb{I} \) is the indicator function that returns 1 if \( y_i \) is among the Top-k predictions, and 0 otherwise. \rev{We apply the Wilcoxon signed-rank test against the best-performing baseline in Top-1 position to check the significance in the performance difference ($p < 0.01$).} 

\section{Results}
%In this section, we discuss our evaluation results and address the research questions.
\label{sec:experimental-results}

\subsection{RQ1: Can TriagerX outperform the baselines?}
\label{sec:eval-contribution-scoring}
\textbf{Approach.} Evaluating our end-to-end framework on literature datasets is challenging due to the lack of detailed interaction data (assignments, commits, pull requests, temporal data, etc.). Hence, we created our own dataset. We compare the Top-k accuracy against the baselines.

% Table \ref{table:triagerx-contrib-scoring} shows that our framework significantly outperforms all baselines across top-\emph{k} metrics for \( k \in \{1, 3, 5, 10, 20\} \). For instance, TriagerX achieves nearly 54\% higher top-1 accuracy than LBT-P and 58\% higher than the closest baseline, RoBERTa-large with CNN classifiers, despite RoBERTa-large having 33\% more parameters on the Openj9 dataset. CodeBERT performed worst for both Openj9 and TS datasets. While the large PLMs showed varying performance on both datasets, TF-IDF-based SVM, however, showed commendable results on both datasets. On the TS dataset, TF-IDF based approach surpassed LBT-P and was only behind TriagerX and RoBERTa based approaches. We observed that TF-IDF worked better on smaller datasets. TriagerX framework performed consistently and outperformed all other approaches in both datasets. 
\noindent\textbf{Results.} Table \ref{table:triagerx-contrib-scoring} shows that our framework consistently outperforms all baselines across Top-k metrics for \( k \in \{1, 3, 5, 10, 20\} \) with statistical significance except for the Top-20 position in the TS dataset. On the OpenJ9 dataset, TriagerX achieves 54\% higher Top-1 accuracy than LBT-P and 58\% higher than the closest large PLM-based baseline, RoBERTa-large with CNN classifiers. CodeBERT showed the lowest performance across both datasets, while TF-IDF-based SVM achieved competitive results, just behind LBT-P in Top-1 accuracy on the TS dataset and ranking just behind TriagerX and RoBERTa-based models. However, TF-IDF's Top-k accuracy declines drastically for $k > 1$ compared to PLM-based approaches.
Overall, TriagerX consistently outperformed all models on both datasets. Optimizing hyperparameters for the framework for smaller \( k \) values, such as Top-1 or 3, using our grid search technique (see Section \ref{sec:ibr-optimization}) leads to improvements for larger \( k \) values as well.
% Overall, TriagerX outperformed all other models consistently on both datasets. When we optimize the framework for smaller \( k \) values—by selecting hyperparameters for IBR to maximize a specific \( k \) value, such as Top-1 or Top-3—using our predefined grid search technique, the improvements typically extend to larger \( k \) values as well. In practice, users also prioritize the accuracy of the highest-ranked recommendations, making precise top recommendations more valuable. 
Despite TriagerX's superior performance compared to other approaches, it still has a low Top-1 accuracy. To understand the misclassifications, we examined the examples and identified dataset imbalance as a key issue, with some developers having far more contributions than others. This bias leads the model to favor more active developers, resulting in poorer performance for those with fewer contributions. Another common reason is when the IBR interaction score fails to complement the misclassification by CBR. For instance, in OpenJ9 issue \href{https://github.com/eclipse-openj9/openj9/issues/17738}{\#17738}, \texttt{@JasonFengJ9} is the actual owner, but CBR predicted other developers who work on similar issues. Manual investigation revealed that similar issues used to calculate interaction scores had few interactions from \texttt{@JasonFengJ9}, thus failing to correct the final recommendation.

\begin{summarybox}
\small\textbf{RQ1 Summary:} The TriagerX framework outperforms the LBT-P baseline by up to 54\% in Top-1 accuracy and also surpasses larger PLM-based approaches consistently.
\end{summarybox}

\vspace{-1em}
\subsection{RQ2: How effective are each of the ranker components?}
\label{sec:eval-cbr}
\textbf{Approach.} We compare the Top-k accuracy of TriagerX IBR and CBR, evaluated separately.
%, against other approaches on five datasets for CBR and two datasets for IBR.
\begin{table}[t]
\small
% \vspace{-1em}
\caption{\revth{Top-k accuracy of TriagerX framework compared to all baselines. P-values are calculated against the best-performing baseline (\ding{171}) where `*' indicates $p<0.01$.}}
\centering
\resizebox{\linewidth}{!}{
\begin{threeparttable}
\begin{tabular}{llccccc}
\toprule
\textbf{Dataset} & \textbf{Method} & \textbf{Top-1} & \textbf{Top-3} & \textbf{Top-5} & \textbf{Top-10} & \textbf{Top-20} \\
\midrule
\multirow{15}{0.6cm}{\centering Openj9} 
& \textbf{TriagerX (CBR+IBR)} & \textbf{0.327*} & \textbf{0.533*} & \textbf{0.633*} & \textbf{0.807*} & \textbf{0.918*} \\
& TriagerX CBR & 0.272 & 0.476 & 0.601 & 0.780 & 0.901 \\
& TriagerX IBR & 0.284 & 0.488 & 0.585 & 0.699 & 0.860 \\
\cmidrule(lr){2-7}
% & \multicolumn{6}{c}{\textbf{Literature Baselines}} \\
\cmidrule(lr){2-7}
% & \multicolumn{6}{c}{\textbf{L PLMs}} \\
\cmidrule(lr){2-7}
& DeBERTa-Large (FCN) & 0.178 & 0.418 & 0.547 & 0.698 & 0.897 \\
& RoBERTa-Large (FCN) & 0.191 & 0.418 & 0.586 & 0.743 & 0.890 \\
& BERT-Large (FCN) & 0.168 & 0.393 & 0.507 & 0.694 & 0.857 \\
& CodeBERT (FCN) & 0.129 & 0.331 & 0.476 & 0.689 & 0.849 \\
& DeBERTa-Large (CNN) & 0.170 & 0.374 & 0.503 & 0.675 & 0.853 \\
& RoBERTa-Large (CNN) & 0.206 & 0.403 & 0.531 & 0.670 & 0.822 \\
& BERT-Large (CNN) & 0.181 & 0.323 & 0.445 & 0.652 & 0.839 \\
& CodeBERT (CNN) & 0.100 & 0.253 & 0.409 & 0.595 & 0.805 \\
\cmidrule(lr){2-7}
& LBT-P \ding{171} & 0.211 & 0.407 & 0.501 & 0.631 & 0.797 \\
& DBRNN-A & 0.127 & 0.300 & 0.454 & 0.627 & 0.775 \\
& MDN & 0.100 & 0.349 & 0.422 & 0.606 & 0.746 \\ 
& TF-IDF + SVM & 0.189 & 0.357 & 0.484 & 0.665 & 0.828 \\ 
\cmidrule(lr){2-7}
& P-value & 1.4e-08 & 3.4e-05 & 1.8e-07 & 1.8e-09 & 3.0e-13 \\
\midrule
\multirow{15}{0.8cm}{\centering TS} 
& \textbf{TriagerX (CBR+IBR)} & \textbf{0.353*} & \textbf{0.615*} & \textbf{0.711*} & \textbf{0.830*} & \textbf{0.930} \\
& TriagerX CBR & 0.324 & 0.582 & 0.682 & 0.812 & 0.920 \\
& TriagerX IBR &  0.278 & 0.487 & 0.564 & 0.650 & 0.720 \\
\cmidrule(lr){2-7}
% & \multicolumn{6}{c}{\textbf{Literature Baselines}} \\
\cmidrule(lr){2-7}
% & \multicolumn{6}{c}{\textbf{L PLMs}} \\
\cmidrule(lr){2-7}
& DeBERTa-Large (FCN) & 0.264 & 0.509 & 0.618 & 0.794 & 0.924 \\
& RoBERTa-Large (FCN) \ding{171} & 0.319 & 0.552 & 0.669 & 0.824 & 0.929 \\
& BERT-Large (FCN) & 0.253 & 0.481 & 0.614 & 0.784 & 0.906 \\
& CodeBERT (FCN) & 0.120 & 0.309 & 0.458 & 0.733 & 0.915 \\
& DeBERTa-Large (CNN) & 0.212 & 0.428 & 0.580 & 0.765 & 0.918 \\
& RoBERTa-Large (CNN) & 0.294 & 0.495 & 0.602 & 0.739 & 0.876 \\
& BERT-Large (CNN) & 0.151 & 0.345 & 0.502 & 0.705 & 0.893 \\
& CodeBERT (CNN) & 0.143 & 0.352 & 0.506 & 0.704 & 0.890 \\
\cmidrule(lr){2-7}
& LBT-P & 0.279 & 0.503 & 0.627 & 0.781 & 0.908 \\
& DBRNN-A & 0.231 & 0.447 & 0.579 & 0.729 & 0.838 \\
& MDN & 0.075 & 0.100 & 0.275 & 0.475 & 0.525 \\
& TF-IDF + SVM & 0.272 & 0.428 & 0.493 & 0.663 & 0.830 \\
\cmidrule(lr){2-7}
& P-value & 8.7e-11 & 1.4e-04 & 4.2e-03 & 4.4e-03 & 1.2e-01 \\
\bottomrule
\end{tabular}
\end{threeparttable}
}
\label{table:triagerx-contrib-scoring}
% \vspace{-2.5em}
\end{table}

\begin{table}[t]
\small
\caption{\revth{Top-k accuracy of all considered baselines on different datasets compared to TriagerX CBR. P-values are calculated against the best-performing baseline (\ding{171}) where `*' indicates $p<0.01$.}}
\centering
\resizebox{\linewidth}{!}{
\begin{tabular}{llccccc}
\toprule
\textbf{Dataset} & \textbf{Method} & \textbf{Top-1} & \textbf{Top-3} & \textbf{Top-5} & \textbf{Top-10} & \textbf{Top-20} \\
\midrule
\multirow{12}{1.5cm}{\centering Google\\Chromium} 
& \textbf{TriagerX CBR} & \textbf{0.345*} & \textbf{0.537*} & \textbf{0.612*} & \textbf{0.710*} & \textbf{0.803*}\\
\cmidrule(lr){2-7}
& DeBERTa-Large (FCN) & 0.285 & 0.474 & 0.567 & 0.677 & 0.767 \\
& RoBERTa-Large (FCN) & 0.267 & 0.461 & 0.551 & 0.660 & 0.755 \\
& BERT-Large (FCN) & 0.255 & 0.433 & 0.520 & 0.630 & 0.715 \\
& CodeBERT (FCN) & 0.224 & 0.403 & 0.493 & 0.606 & 0.697 \\
& DeBERTa-Large (CNN) & 0.251 & 0.432 & 0.525 & 0.639 & 0.738 \\
& RoBERTa-Large (CNN) & 0.281 & 0.475 & 0.564 & 0.671 & 0.763 \\
& BERT-Large (CNN) & 0.271 & 0.455 & 0.549 & 0.655 & 0.743\\
& CodeBERT (CNN) & 0.159 & 0.319 & 0.399 & 0.519 & 0.634 \\
\cmidrule(lr){2-7}
& LBT-P \ding{171} & 0.318 & 0.499 & 0.578 & 0.676 & 0.763 \\
& DBRNN-A & 0.183 & 0.318 & 0.385 & 0.482 & 0.581 \\
& TF-IDF + SVM & 0.204 & 0.310 & 0.376 & 0.454 & 0.529 \\
\cmidrule(lr){2-7}
& P-value & 4.4e-18 & 5.7e-18 & 4.0e-18 & 5.5e-18 & 5.4e-18  \\
\midrule
\multirow{12}{1.5cm}{\centering Mozilla\\Core} 
& \textbf{TriagerX CBR} & \textbf{0.340*} & \textbf{0.521*} & \textbf{0.598*} & \textbf{0.700*} & \textbf{0.805*} \\
\cmidrule(lr){2-7}
& DeBERTa-Large (FCN) & 0.257 & 0.437 & 0.521 & 0.639 & 0.744 \\
& RoBERTa-Large (FCN) & 0.276 & 0.458 & 0.540 & 0.650 & 0.749 \\
& BERT-Large (FCN) & 0.215 & 0.378 & 0.461 & 0.571 & 0.681 \\
& CodeBERT (FCN) & 0.206 & 0.371 & 0.455 & 0.570 & 0.678 \\
& DeBERTa-Large (CNN) & 0.269 & 0.445 & 0.533 & 0.639 & 0.737 \\
& RoBERTa-Large (CNN) \ding{171} & 0.306 & 0.490 & 0.568 & 0.668 & 0.758 \\
& BERT-Large (CNN) & 0.258 & 0.432 & 0.514 & 0.624 & 0.725 \\
& CodeBERT (CNN) & 0.268 & 0.447 & 0.532 & 0.640 & 0.739 \\
\cmidrule(lr){2-7}
& LBT-P & 0.279 & 0.471 & 0.553 & 0.655 & 0.748 \\
& DBRNN-A & 0.164 & 0.290 & 0.367 & 0.481 & 0.594\\
& TF-IDF + SVM & 0.238 & 0.386 & 0.454 & 0.546 & 0.638 \\
\cmidrule(lr){2-7}
& P-value & 1.1e-07 & 2.8e-06 & 1.9e-04 & 3.5e-11 & 7.2e-07 \\
\midrule
\multirow{12}{1.5cm}{\centering Mozilla\\Firefox} 
& \textbf{TriagerX CBR} & \textbf{0.272*} & \textbf{0.471*} & \textbf{0.576*} & \textbf{0.718*} & \textbf{0.835*} \\
\cmidrule(lr){2-7}
& DeBERTa-Large (FCN) & 0.221 & 0.402 & 0.488 & 0.646 & 0.801 \\
& RoBERTa-Large (FCN) & 0.218 & 0.400 & 0.505 & 0.642 & 0.781 \\
& BERT-Large (FCN) & 0.213 & 0.353 & 0.445 & 0.585 & 0.748 \\
& CodeBERT (FCN) & 0.193 & 0.366 & 0.454 & 0.597 & 0.760 \\
& DeBERTa-Large (CNN) & 0.199 & 0.368 & 0.473 & 0.627 & 0.798 \\
& RoBERTa-Large (CNN) \ding{171} & 0.248 & 0.441 & 0.534 & 0.671 & 0.801\\
& BERT-Large (CNN) & 0.148 & 0.306 & 0.389 & 0.516 & 0.670 \\
& CodeBERT (CNN) & 0.219 & 0.385 & 0.483 & 0.619 & 0.754 \\
\cmidrule(lr){2-7}
& LBT-P & 0.243 & 0.423 & 0.524 & 0.646 & 0.788 \\
& DBRNN-A & 0.135 & 0.253 & 0.334 & 0.441 & 0.612 \\
& TF-IDF + SVM & 0.221 & 0.388 & 0.454 & 0.540 & 0.623 \\
\cmidrule(lr){2-7}
& P-value & 2.3e-04 & 2.6e-03 & 4.2e-03 & 8.7e-03 & 1.0e-04 \\
\bottomrule
\end{tabular}
}
\label{table:model-comparison}
\end{table}

\noindent\textbf{Effectiveness of TriagerX IBR.} We evaluate TriagerX IBR on partner datasets, as literature datasets lack developer interaction data. Table \ref{table:triagerx-contrib-scoring} shows that IBR, using similarity-based retrieval, outperforms other methods in Top-\{1,3\} accuracy on the Openj9 dataset and performs competitively on the TS dataset signifying the effectiveness of using developer interaction data for bug triaging. In contrast, MDN underperforms on both datasets due to its reliance on unigram models that focus on term frequency rather than contextual meaning. For example, MDN may link a ``VM crash" issue to a ``UI crash" rather than a more relevant ``JVM failure." Additionally, the approach weighs developer contributions equally regardless of time which may prioritize developers who frequently contributed to similar issues long ago than someone who recently contributed to similar issues. Our approach overcomes these limitations by using a context-sensitive PLM and time-based contribution scoring. However, integrating both IBR and CBR delivers the best overall performance across both datasets. We analyzed misclassifications for IBR and found that team leads or project managers often engage with issues through discussions or mentions without being the actual solvers. Despite assigning low scores to discussions, these interactions can still lead to inaccurate recommendations. For instance, in OpenJ9 Issue \href{https://github.com/eclipse-openj9/openj9/issues/19197}{\#19197}, user \texttt{@pshipton} frequently participated in discussion but was not the fixer. The IBR algorithm might incorrectly identify them as a potential resolver for similar issues, slightly lowering Top-1 accuracy.

\noindent\textbf{Effectiveness of TriagerX CBR.} Since TriagerX CBR relies solely on the textual content of bug reports, we had a broader range of evaluation options. Table \ref{table:triagerx-contrib-scoring} presents the evaluation results of CBR for our datasets, and Table \ref{table:model-comparison} provides a comparative analysis with literature datasets. CBR consistently outperforms the literature baselines like LBT-P and DBRNN-A. CBR also achieves higher Top-k accuracy compared to larger PLMs with naive-FCN or CNN classifiers. DeBERTa performed better in the GC dataset, while RoBERTa in the MC dataset. However, TF-IDF performed significantly worse on larger datasets like GC and MC, likely due to its reliance on sparse feature representations. We also analyzed cases where the CBR model failed to predict the correct developer, often due to data imbalance. Additionally, the model sometimes makes incorrect recommendations when multiple developers work on similar issues. Although different developers might resolve the issue, these predictions are considered incorrect because the datasets assign each bug to only one developer. For example, consider the following Openj9 issues \href{https://github.com/eclipse-openj9/openj9/issues/18871}{\#18871} and \href{https://github.com/eclipse-openj9/openj9/issues/18873}{\#18873} that are extremely similar in title and with the same component labels but were fixed by different developers.

\noindent\textbf{\revth{Analysis of IBR Contributions.}} \revth{We further analyze OpenJ9 issues where CBR fails at Top-1, but the combined TriagerX framework (CBR+IBR) correctly predicts the resolver at Top-1. We group these 43 corrected cases using the repository’s developer-assigned labels for further analysis. Table~\ref{tab:ibr_correction_groups} summarizes the prevalence of each issue type. The total label count exceeds 43 as issues are multi-labeled in OpenJ9 (e.g., a \textit{test failure} in a \textit{VM component}).}

\revth{We find that over 60\% of the issues corrected by IBR integration are test-related. These reports are typically log-heavy and template-like (e.g., stack traces and test harness output), which obscures ownership cues and causes CBR to favor developers associated with broadly similar test failures. IBR mitigates this by reordering candidates using interaction signals from similar test issues. For example, Issues \href{https://github.com/eclipse-openj9/openj9/issues/17780}{\#17780}
 and \href{https://github.com/eclipse-openj9/openj9/issues/18384}{\#18384}
 exhibit the same failure mode \textit{(cmdLineTester\_criu\_nonPortableRestore)} where CBR ranked \texttt{@dsouzai} first, while the true resolver \texttt{@JasonFengJ9} appeared lower. IBR scores enabled TriagerX to promote \texttt{@JasonFengJ9} to Top-1 in both cases based on recent interactions with similar related issues.}

\revth{Core Runtime issues form the second-largest group among the cases corrected by IBR. The main difficulty in these cases is that runtime components in OpenJ9 are maintained by multiple developers over time, and many runtime bug reports are textually similar to past issues that were resolved by different developers at different points in the project’s history. This makes it challenging for a content-only ranker to distinguish the most appropriate resolver based on textual similarity alone. This behavior is evident in our results--21 out of 22 runtime issues (95.5\%) corrected by IBR, the true resolver already appears in CBR’s Top-5 predictions. Thus, CBR generally retrieves a relevant shortlist but often misorders candidates at rank~1 when several developers have historically resolved similar issues within the same subsystem. Analysis of OpenJ9 resolution history confirms that runtime labels reflect shared maintenance rather than exclusive ownership. For example, issues labeled \texttt{comp:jvmti} have been resolved by multiple developers, including \texttt{@babsingh} as well as \texttt{@llxia} and \texttt{@thallium}. When a new \texttt{jvmti} report closely resembles past reports handled by different maintainers, CBR may reasonably favor a frequent—but not \textit{currently} most relevant resolver. For example, in issues \href{https://github.com/eclipse-openj9/openj9/issues/17871}{\#17871}
 and \href{https://github.com/eclipse-openj9/openj9/issues/18730}{\#18730}, CBR ranks \texttt{@llxia} or \texttt{@thallium} at Top-1 while placing the true resolver \texttt{@babsingh} lower. IBR corrects these errors by using interaction-based evidence, using recent interactions to re-rank candidates and promote \texttt{@babsingh} to Top-1.}
 
 \revth{Finally, Platform Infrastructure issues include system and security-related components where issues are often resolved by the developer who is actively handling the affected infrastructure or security configuration at the time, rather than by a stable component maintainer. Consequently, textual similarity provides weak ownership signals, and CBR frequently prioritizes historically frequent contributors instead of the currently responsible developer. IBR corrects these cases by leveraging recent interaction evidence, allowing TriagerX to recover the correct resolver in 12 cases (e.g., Issues \href{https://github.com/eclipse-openj9/openj9/issues/18436}{\#18436}, \href{https://github.com/eclipse-openj9/openj9/issues/19049}{\#19049}).}

\begin{table}[t]
\centering
\caption{\revth{Issue types for which IBR corrected an incorrect Top-1 prediction from CBR. Issue types are non-mutually exclusive; an issue may belong to multiple types.}}
\label{tab:ibr_correction_groups}
\resizebox{\linewidth}{!}{%
\begin{tabular}{lp{3cm}cc}
\toprule
\textbf{Issue Type} & \textbf{Associated Labels} & \textbf{\# Issues} & \textbf{Percentage} \\
\midrule
Test Issues 
& test failure, test excluded, \newline comp:test 
& 29 & 67.4\% \\
\midrule
Core Runtime 
& comp:vm, comp:jit, \newline comp:jvmti, comp:ras 
& 22 & 51.2\% \\
\midrule
Platform Infrastructure 
& comp:infra, comp:build, \newline comp:openssl, comp:fips 
& 12 & 27.9\% \\
\bottomrule
\end{tabular}
}
\end{table}

\begin{summarybox}
\small\textbf{RQ2 Summary:} TriagerX CBR consistently outperforms baseline models like LBT-P and DBRNN-A, as well as larger PLMs, achieving superior Top-k accuracy across all datasets. TriagerX IBR also shows strong performance, particularly when used in conjunction with CBR within the TriagerX framework.
\end{summarybox}
\vspace{-1em}
\subsection{RQ3: How do multiple PLMs influence CBR accuracy?}
\label{sec:rq-multiple-plm}
\label{sec:rq-multiple-plm-overall}
\textbf{Approach.} \revth{We assess how an ensemble of multiple PLMs influences the accuracy of CBR by directly comparing it with the individual PLMs used to construct it. Concretely, we train each PLM independently using the same architecture and the same number of classifiers (i.e., 3 CNN classifiers) as in the full CBR and compare the results.} We answer two sub-RQs:

\textbf{RQ3.1} Overall Performance: We evaluate the performance of TriagerX CBR with individual PLMs and in ensemble to determine the impact of using multiple PLMs.

\textbf{RQ3.2} Class-wise Performance: We analyze how embedding diversity from different PLMs affects class-wise Top-1 accuracy, comparing TriagerX CBR’s performance with the PLM ensemble against individual PLMs on specific classes from a benchmark dataset.

\begin{figure}
    \centering
    \includegraphics[width=2.25in]{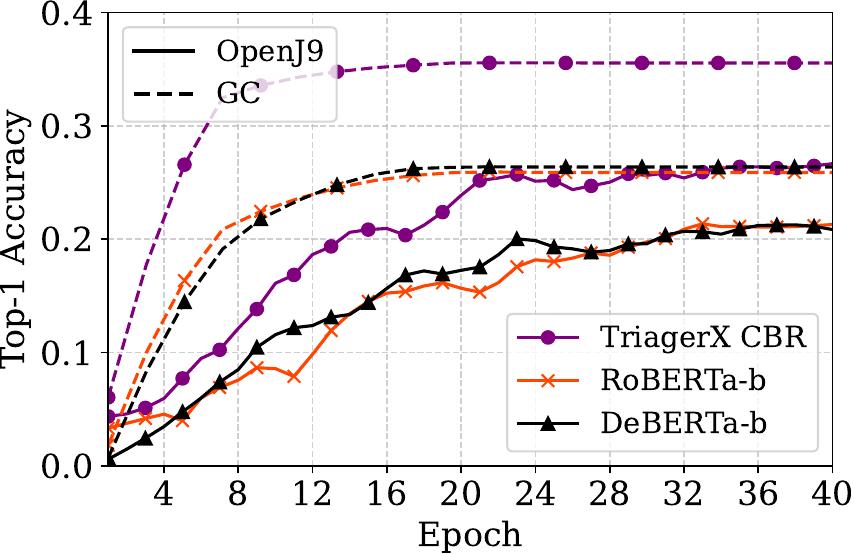}
    \caption{Top-1 test accuracy of TriagerX CBR model compared to the RoBERTa and DeBERTa base models when trained separately with 3 classifiers on different datasets.}
    \label{fig:top-1-accuracy-openj9}
    % \vspace{-em}
\end{figure}

\noindent\textbf{Overall Performance.}
The test accuracy results shown in Figure \ref{fig:top-1-accuracy-openj9} indicate that both PLMs, despite being pretrained on different corpora, achieve similar accuracy when fine-tuned separately on the same dataset. However, their ensemble within CBR outperforms each model, due to the complementary knowledge each PLM brings from its unique pertaining. A similar trend is observed in both the OpenJ9 and GC datasets, despite their differences in size and textual content, with accuracy improving significantly when base PLMs are ensembled. For consistency and fair evaluation, all base models were trained with identical hyperparameters, including frozen encoder layers and the number of classifiers. This confirms the effectiveness of combining multiple smaller PLMs, as the ensemble approach leverages diverse feature representations and thus enhances the robustness and generalizability of TriagerX CBR’s predictions.
\begin{figure}[t]
\centering
\includegraphics[width=\linewidth]{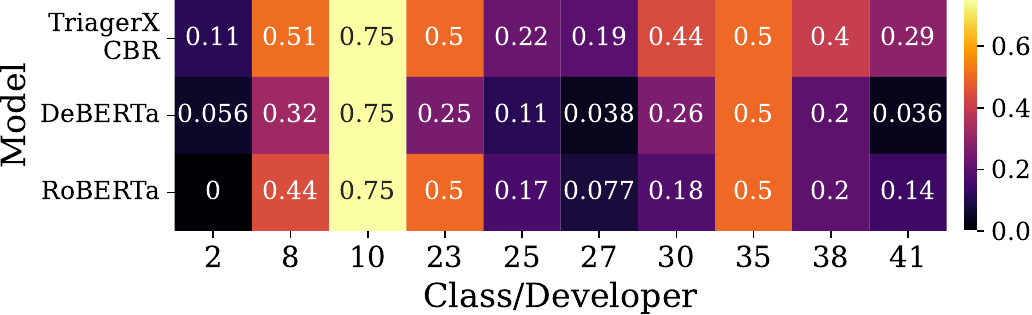}
    \caption{Class-wise Top-1 accuracy on the OpenJ9 dataset.}
    \label{fig:openj9-top1-classwise}
    % \vspace{-1em}
\end{figure}

% \begin{wrapfigure}[10]{r}{0.55\linewidth}
%     % \vspace{-2em}
%     \includegraphics[width=0.9\linewidth]{figures/openj9-top1-classwise.pdf}
%     \caption{Class-wise Top-1 accuracy on the OpenJ9 dataset.}
%     \label{fig:openj9-top1-classwise}
% \end{wrapfigure}

\noindent\textbf{Class-wise Performance.}
We discussed that combining multiple PLMs enhances overall performance. Here, we examine if this improvement is due to the orthogonality among models. We quantify orthogonality for PLM \(P_i\) as: 
\begin{equation}
     O(P_i) = |C_{P_i} - \left(\bigcup_{j \neq i} C_{P_j}\right)|
 \end{equation}
where \(C_{P_i}\) is the set of bug reports uniquely triaged by \(P_i\). Higher \(O(P_i)\) indicates that \(P_i\) handles unique cases well, suggesting benefits from combining PLMs.  If there is a high degree of orthogonality among the base PLMs, combining them should improve class-wise Top-1 accuracy across the dataset. We compare the accuracy of TriagerX CBR using both DeBERTa and RoBERTa in ensemble versus individually on the OpenJ9 dataset. Figure \ref{fig:openj9-top1-classwise} shows that using multiple PLMs improves Top-1 accuracy across nearly all selected classes. However, this improvement is not always linear—some classes benefit significantly (e.g., Class 8, 30, 38), while others see little to no change (e.g., Class 10, 25). \revtwo{This variation arises because the ensemble fuses embeddings from multiple PLMs rather than simply combining predictions. For classes where both PLMs already perform well, or where both struggle similarly (e.g., very few training data), additional embeddings may not provide further gains. Conversely, for classes where PLMs exhibit complementary strengths, the fused embeddings provide richer latent features that allow the classifier to learn more discriminative representations \cite{lester2020multiple, tsukagoshi2022comparison}. Therefore, although some classes show little change, the consistent improvement at the aggregate level demonstrates that the PLMs capture complementary patterns that benefit the overall system.}

% This variation occurs because the ensemble method does not merely rely on individual predictions but fuses the embeddings from different PLMs which is then used by the classifier. Even when a single PLM struggles with a particular class, its embeddings may still provide useful latent features that, when combined with another PLM allows the classifier to learn a better representation \cite{lester2020multiple, tsukagoshi2022comparison}.

% For example, in Class 2, RoBERTa alone fails entirely (0 accuracy), and DeBERTa achieves minimal accuracy (0.056). Yet, the ensemble achieves a modest improvement (0.11). This suggests that although RoBERTa fails to classify Class 2 directly, its embeddings still contribute complementary representational features that help refine the final classification when concatenated with DeBERTa’s embeddings within TriagerX CBR.

\begin{summarybox}
\small\textbf{RQ3 Summary:} Combining multiple PLMs in TriagerX CBR enhances overall and class-wise accuracy. The improved performance is attributed to the orthogonality between models, resulting in higher Top-1 accuracy and better class-specific results compared to individual PLMs.
\end{summarybox}
% \vspace{-1em}

\subsection{RQ4: Do the design decisions impact CBR performance?}
\label{sec:rq4-cbr-design}
% \begin{wrapfigure}[11]{r}{0.45\linewidth}  % Reduce width to 0.4
%     \centering
%     \vspace{-1.5em}
% \includegraphics[width=0.9\linewidth]{figures/cnn-vs-fcn.pdf}
%     \caption{Comparison of CNN vs. vanilla FCN-based classifiers.}
%     \label{fig:cnn-vs-fcn}
% \end{wrapfigure}

\textbf{Approach.} We investigated three design choices affecting CBR performance: \rev{(1) Different PLM combinations}, (2) replacing the CNN-based classifier with a vanilla FCN classifier, and (3) varying the number of classifiers in the TriagerX CBR model.

\noindent\textbf{PLM Combinations.} \rev{We evaluate multiple PLM combinations on 2 datasets. Table \ref{table:plm-comparison} shows that all combinations yield comparable results, supporting the general robustness of our ensemble design. Among them, the RoBERTa- and DeBERTa-base pair consistently achieved slightly better performance, and thus, we adopt them as the default encoders in TriagerX.}

\begin{table}[t]
\small
\caption{Accuracy of different PLM combinations in CBR.}
\centering
\resizebox{\linewidth}{!}{
\begin{tabular}{llccccc}
\toprule
\textbf{Dataset} & \textbf{Method} & \textbf{Top-1} & \textbf{Top-3} & \textbf{Top-5} & \textbf{Top-10} & \textbf{Top-20} \\
\midrule
\multirow{6}{1.5cm}{\centering Google\\Chromium} 
& \textbf{RoBERTa + DeBERTa} & \textbf{0.345} & \textbf{0.537} & 0.612 & \textbf{0.710} & \textbf{0.803}\\
& RoBERTa + BERT & 0.337 & 0.522 & 0.601 & 0.699 & 0.784 \\
& DeBERTa + BERT & 0.343 & 0.533 & \textbf{0.613} & 0.701 & 0.795 \\
& CodeBERT + BERT & 0.335 & 0.529 & 0.604 & 0.700 & 0.791 \\
& CodeBERT + RoBERTa & 0.342 & 0.529 & 0.611 & 0.703 & 0.791 \\
& CodeBERT + DeBERTa & 0.341 & 0.533 & 0.602 & 0.702 & 0.790 \\
\midrule
\multirow{6}{1.5cm}{\centering Openj9} 
& \textbf{RoBERTa + DeBERTa} & \textbf{0.272} & 0.476 &\textbf{ 0.601} & \textbf{0.780} & \textbf{0.901} \\
& RoBERTa + BERT & 0.251 & \textbf{0.482} & 0.596 & 0.756 & 0.887 \\
& DeBERTa + BERT & 0.264 & 0.476 & 0.590 & 0.746 & 0.899 \\
& CodeBERT + BERT & 0.261 & 0.463 & 0.580 & 0.753 & 0.883 \\
& CodeBERT + RoBERTa & 0.246 & 0.461 & 0.588 & 0.766 & 0.893 \\
& CodeBERT + DeBERTa & 0.259 & 0.481 & 0.588 & 0.748 & 0.891 \\
\bottomrule
\end{tabular}
}
\label{table:plm-comparison}
% \vspace{-1em}
\end{table}

\begin{figure}[t]
\centering
\includegraphics[width=2.25in]{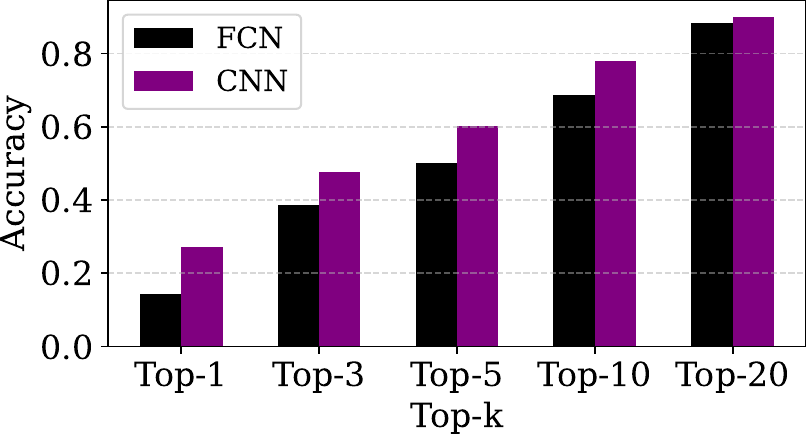}
    \caption{Comparison of CNN vs. vanilla FCN-based classifiers.}
    \label{fig:cnn-vs-fcn}
\end{figure}

\begin{figure}[t]
\centering
\includegraphics[width=2.25in]{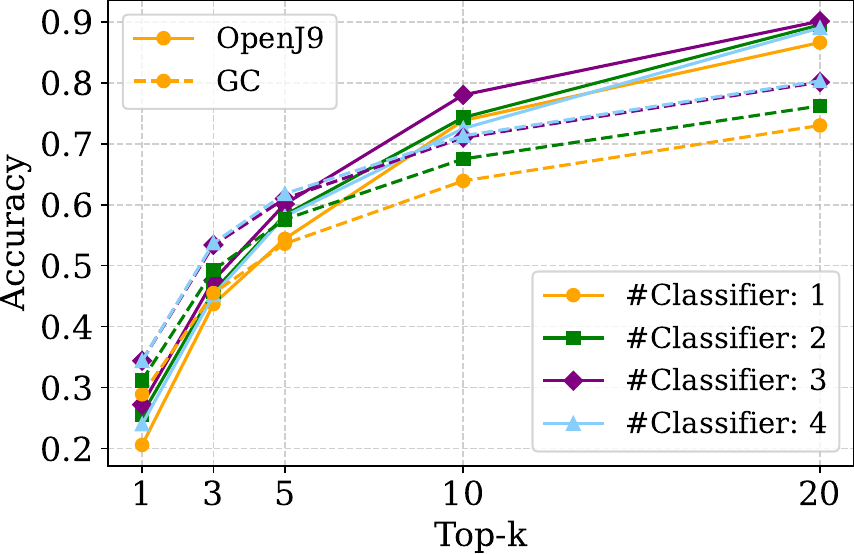}
    \caption{Impact of the number of classifiers on Top-k accuracy across OpenJ9 and Google Chromium (GC) datasets.}
    \label{fig:classifiers-vs-topk}
    % \vspace{-1em}
\end{figure}

\noindent\textbf{CNN vs. Vanilla FCN.} Replacing the CNN-based classifier with a vanilla FCN in the TriagerX CBR component showed that the CNN outperforms the FCN in all Top-k metrics on the Openj9 dataset, with more than 80\% improvement in Top-1 accuracy. CNN's ability to capture contextual patterns from transformer hidden states contributed to its superior performance, as shown in Figure \ref{fig:cnn-vs-fcn}.

\noindent\textbf{Optimal Number of Classifiers.} We evaluated the effect of varying the number of classifiers in TriagerX CBR. As shown in Figure \ref{fig:classifiers-vs-topk}, performance generally improves with an increasing number of classifiers. For smaller datasets like OpenJ9, using more than 3 classifiers leads to overfitting and degraded performance. In contrast, for larger datasets like GC, increasing the number of classifiers beyond 3 yields only marginal improvement, with nearly identical Top-k accuracy curves for 3 and 4 classifiers. More classifiers also increase trainable parameters and training times. Therefore, we selected 3 classifiers for CBR across all datasets.

\begin{summarybox}
\small\textbf{RQ4 Summary:} The ensemble technique improves accuracy regardless of the PLM combination, CNN-based classifier consistently outperformed FCN, and using up to three classifiers provided the best balance between performance and generalizability.
\end{summarybox}

% \begin{figure}[htbp]
%     \centering
%     \includegraphics[width=0.8\linewidth]{figures/cnn-vs-fcn.pdf}
%     \caption{Comparison of CNN vs. vanilla FCN-based classifiers.}
%     \label{fig:cnn-vs-fcn}
% \end{figure}

\subsection{RQ5. Do hyperparameter choices affect IBR performance?}

\textbf{Approach.} We examine the impact of hyperparameters and configurations on IBR performance through three analyses: varying the time-decay factor \(\lambda\), adjusting the similarity threshold \(\tau\), and analyzing the choice of rank aggregation (RAgg) algorithm.

\begin{figure}
    \centering
    \includegraphics[width=2.75in]{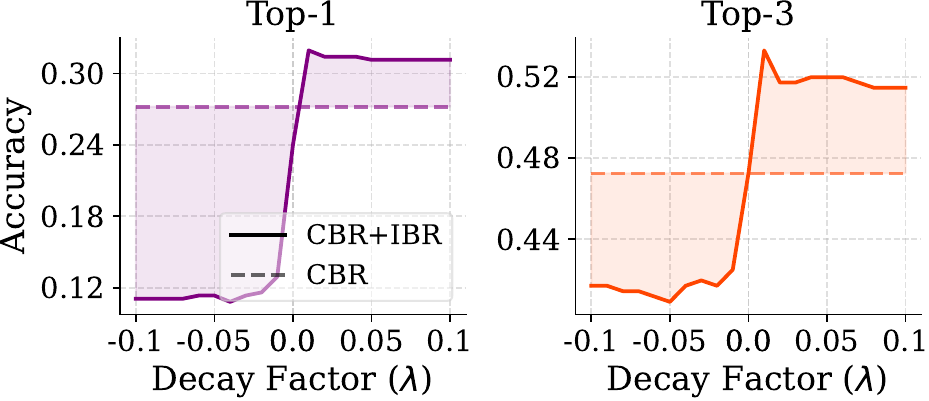}
    \caption{Impact of time-decay factor on OpenJ9 dataset.}
    \label{fig:time-factor}
    % \vspace{-1em}
\end{figure}

\noindent\textbf{Impact of Time-factored Scoring ($\lambda$).} We evaluate the impact of the time-factored scoring parameter \(\lambda\), with negative values prioritizing older contributions and positive values emphasizing recent ones. Figure \ref{fig:time-factor} shows that negative \(\lambda\) values result in poorer performance, with lower Top-{1, 3} accuracy scores. Conversely, positive \(\lambda\) values, such as \(\lambda = 0.01\), significantly enhance accuracy by giving more weight to recent contributions. When \(\lambda = 0\), all contributions are treated equally, leading to similar performance to CBR but less effective than weighting recent contributions more. This asserts the importance of prioritizing recent interactions for better recommendations.

\begin{figure}
    \centering
    \includegraphics[width=2.25in]{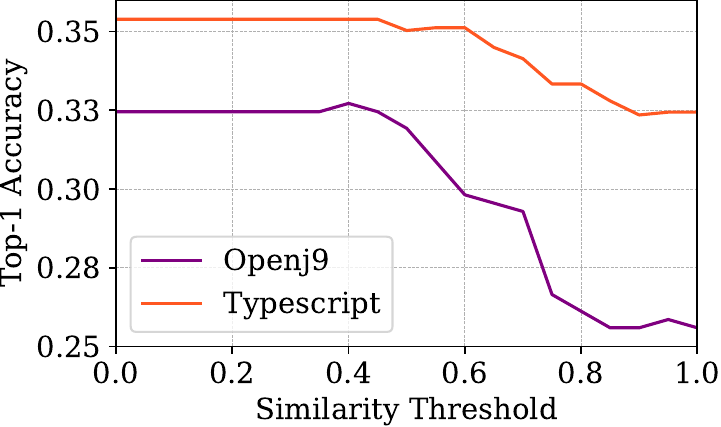}
    \caption{Effect of similarity threshold over Top-1 accuracy.}
    \label{fig:top1-vs-threshold}
    % \vspace{-1em}
\end{figure}

\noindent\textbf{Effect of Similarity Threshold ($\tau$).} We evaluate how varying the similarity threshold affects Top-1 accuracy in TriagerX. A lower threshold generally improves accuracy by including more similar issues, which increases developer interactions. However, very low thresholds increase computational demands and may introduce less relevant issues. Conversely, higher thresholds reduce the number of considered issues, potentially decreasing accuracy. Figure \ref{fig:top1-vs-threshold} indicates that the optimal threshold in our studied datasets lies around 0.40 - 0.45.

\begin{table}
\small
\caption{Comparison of Borda count ranking aggregation compared to ours.}
\centering
\resizebox{\linewidth}{!}{
\begin{tabular}{cc|ccccc}
\toprule
\textbf{Dataset} & \textbf{Aggregation Method} & \textbf{Top-1} & \textbf{Top-3} & \textbf{Top-5} & \textbf{Top-10} & \textbf{Top-20} \\
\midrule
\multirow{2}{*}{\centering Openj9}  & Borda Count & 0.277 & 0.506 & 0.622 & 0.755 & 0.886\\
 & \textbf{WRA} & \textbf{0.327} & \textbf{0.533} & \textbf{0.633} & \textbf{0.807} & \textbf{0.918}  \\
 \midrule
 \multirow{2}{*}{\centering TS}  & Borda Count & \textbf{0.357} & 0.607 & 0.680 & 0.774 & 0.888\\
 & \textbf{WRA} & 0.353 & \textbf{0.615} & \textbf{0.711} & \textbf{0.830} & \textbf{0.930}  \\
\bottomrule
\end{tabular}%
}
\label{table:rank-agg}
% \vspace{-1em}
\end{table}

% \begin{table}
% \small
% \caption{Comparison of Borda ranking aggregation compared to ours.}
% \centering
% \resizebox{8.5cm}{!}{
% \begin{tabular}{cc|ccccc}
% \toprule
% \textbf{Dataset} & \textbf{Aggregation\\Method} & \textbf{K=1} & \textbf{K=3} & \textbf{K=5} & \textbf{K=10} & \textbf{K=20} \\
% \midrule
% penj9 & Borda & 0.240 & 0.452 & 0.581 & 0.725 & 0.890 \\
% %   \textbf{Ours} & \textbf{0.272} & \textbf{0.476} & \textbf{0.601} & \textbf{0.780} & \textbf{0.901} \\
% % \bottomrule
% \end{tabular}
% }
% \label{table:rank-aggregation-comparison}
% \end{table}

\noindent\textbf{Selection of RAgg Algorithm.} We initially used the Borda count method for rank aggregation, which assigns points based on rank positions from different rankers. However, it treats all rankers equally, leading to skewed recommendation scores when developers have limited interaction with similar bugs. To address this, we employed a Weighted Ranking Aggregation (WRA) method as discussed in Section \ref{sec:ragg}. The tunable weight factor \( W_f \) in Equation \ref{eq:rank-agg} optimally balances IBR and CBR rankings, enhancing adaptability across datasets while keeping CBR as the primary ranker. Moreover, WRA does not affect CBR ranking when a developer lacks relevant interaction history, improving accuracy in most cases (Table \ref{table:rank-agg}).

\begin{figure}
    \centering
    \includegraphics[width=0.9\linewidth]{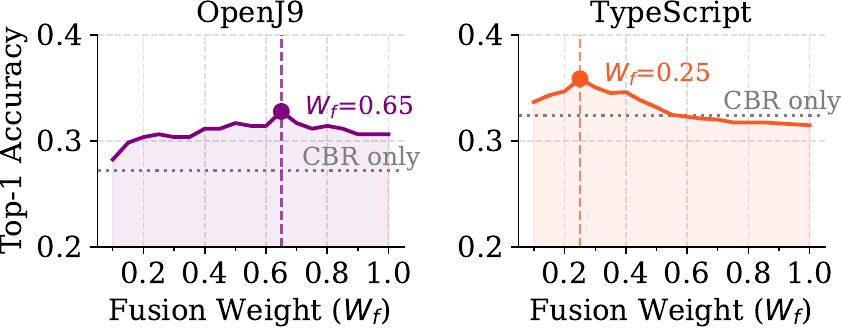}
    \caption{Sensitivity of fusion weight $W_f$ in different datasets.}
    \label{fig:wf_sensitivity}
    % \vspace{-0.5em}
\end{figure}

\noindent\textbf{\revm{Sensitivity of Fusion Weight ($W_f$).}} \revm{Figure~\ref{fig:wf_sensitivity} illustrates the sensitivity of Top-1 accuracy to $W_f$ in Equation \ref{eq:rank-agg} across datasets with interaction history. In OpenJ9, accuracy remains stable across a broad range but peaks at $W_f=0.65$, reflecting the benefit of heavily weighting IBR signals in a repository with dense contributor interactions. In contrast, TypeScript peaks at $W_f=0.25$ and degrades steadily as $W_f$ increases, suggesting that over-relying on interaction signals in sparse contributor settings may negatively impact performance. In both cases, the combined framework consistently outperforms the CBR-only baseline (dotted line), demonstrating that IBR integration is broadly beneficial regardless of the exact weight chosen. For new deployments, repositories with dense interaction histories should favor higher $W_f$ values, while sparse ones should use lower values, with final tuning via grid search (Section~\ref{sec:ibr-optimization}) on a small validation set.}

\begin{summarybox}
\small\textbf{RQ5 Summary:} Optimizing the time-decay factor \( \lambda \) to favor recent contributions, choosing an optimal similarity threshold \( \tau \) for balance between accuracy and computational efficiency, and employing a custom weighted rank aggregation algorithm significantly enhance the performance of TriagerX IBR.
\end{summarybox}

\section{Industrial Deployment of TriagerX}
\label{sec:deployment}
We integrated TriagerX into our industrial partner's dev environment to assess its effectiveness. At this moment, TriagerX serves triaging recommendations to the triagers for the Eclipse OpenJ9 project via GitHub Workflows. The framework runs in a Docker container for efficient deployment and maintenance, delivering recommendations via GitHub bot comments and Slack notifications. 
% This section offers insights of our deployment and on the lessons learned.
% We analyze its impact on the workflow and collect developer feedback to guide future improvements.
%\subsection{Deployment Pipeline}
%TriagerX is integrated into the Eclipse OpenJ9 project via GitHub Workflows. 
% It initially considers 17 active developers for effective recommendations and feedback, and predicts component labels for 9 major components with separate models for developer and component predictions.

\subsection{TriagerX Deployment} 

\noindent\textbf{\revm{Deployment Settings.}} \revm{Before deployment, we followed the same data preparation, model training, and IBR precomputation pipeline described in Sections \ref{sec:datasets}--\ref{sec:ibr-optimization}. The target developer set comprised the 17 most active contributors confirmed by the partner, who collectively resolved the majority of issues at the time of deployment. TriagerX is designed to support incremental retraining, allowing the model to be updated once a new developer accumulates at least 20 resolved issues, ensuring the system remains current as the contributor base evolves. More broadly, the model retraining in the industrial deployment was triggered manually on a monthly basis by the partner team, representing a low-maintenance overhead that was found to be sufficient for keeping recommendations current, given the repository's issue resolution rate.}
\begin{figure}[t]
\centering
\includegraphics[width=0.75\linewidth]{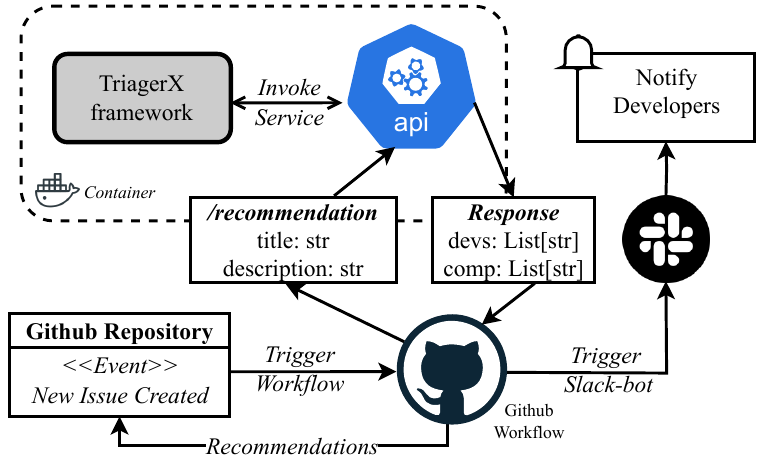}
    \caption{Deployment pipeline of TriagerX framework.}
    \label{fig:triagerx-deployment}
    % \vspace{-1em}
\end{figure}

\noindent\textbf{\revm{Technical Integration.}} \rev{We deployed TriagerX on a CPU-only machine---IBM Power9@2.20 GHz with 8GB memory, as this was the hardware provided by the team. For the OpenJ9 repository, the system operates within a 4 GB memory footprint and produces component and developer recommendations in approximately 3-4 seconds, meeting the team’s real-time responsiveness needs. We also validated performance on an Intel Xeon Gold 5320@2.20 GHz CPU with the same RAM, which showed similar latency. However, when we tested under lab settings with a suitable GPU (NVIDIA V100), inference latency can be reduced to 100–200 ms. While inference is efficient on CPU, training the CBR model still requires a CUDA-enabled GPU to ensure practical training times.}

\begin{figure}[!t]
\centering
\includegraphics[width=0.9\linewidth]{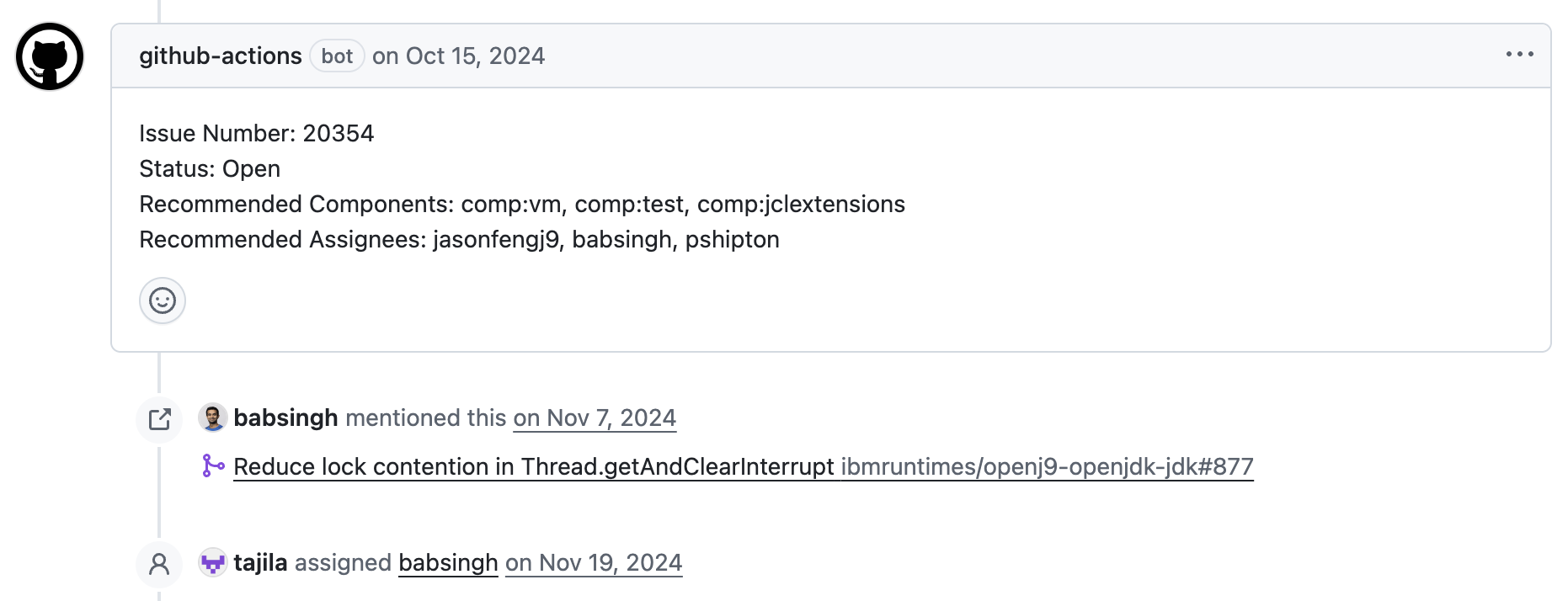}
    \caption{An example (\href{https://github.com/eclipse-openj9/openj9/issues/20354}{\#20354}) of developer recommendation in the OpenJ9 Github repository.}
    \label{fig:example-recommendation}
    % \vspace{-1em}
\end{figure}

\revm{The integration is orchestrated through a GitHub Actions workflow that triggers automatically whenever a new issue is opened in the repository. Upon triggering, the workflow extracts the issue title and body using the GitHub Actions context, then invokes a Node.js script that sends a POST request to the TriagerX REST API endpoint (\texttt{/recommendation}). The API processes the input and returns a ranked list of recommended developers and components, which are then posted directly as a bot comment on the issue page. Figure \ref{fig:triagerx-deployment} illustrates the overall deployment pipeline, while Figure \ref{fig:example-recommendation} shows an example where a recommended developer was subsequently assigned to the issue. The workflow was intentionally designed to be lightweight and non-intrusive; no developer onboarding or formal training sessions were required, as the bot comment format was immediately interpretable without guidance.}

\subsection{GPU Utilization of TriagerX CBR} 
\label{sec:model-efficiency}
One of the requirements from our partner was to be able to train the TriagerX models with as less memory (GPU) as possible. TriagerX CBR has fewer parameters than traditional \textbf{large} PLMs. This size reduction is achieved by combining the \textbf{base} variants of different PLM families, which, when ensembled strategically, result in a lower total parameter count compared to a typical large PLM while maintaining superior accuracy, as discussed in RQ3. For example, RoBERTa and DeBERTa base have 125M and 139M parameters, respectively. When combined in CBR, they total around 271M, which is fewer than DeBERTa-Large's 405M parameters alone, excluding classifiers (see Appendix \ref{app:plm-variants}). The improvement after combining the PLMs arises from the orthogonal and complementary characteristics of each model, similar to the boosting technique in AdaBoost \cite{schapire2013explaining}, where weak learners are aggregated to strengthen predictions.
\begin{figure}[!t]
\centering
\includegraphics[width=0.9\linewidth]{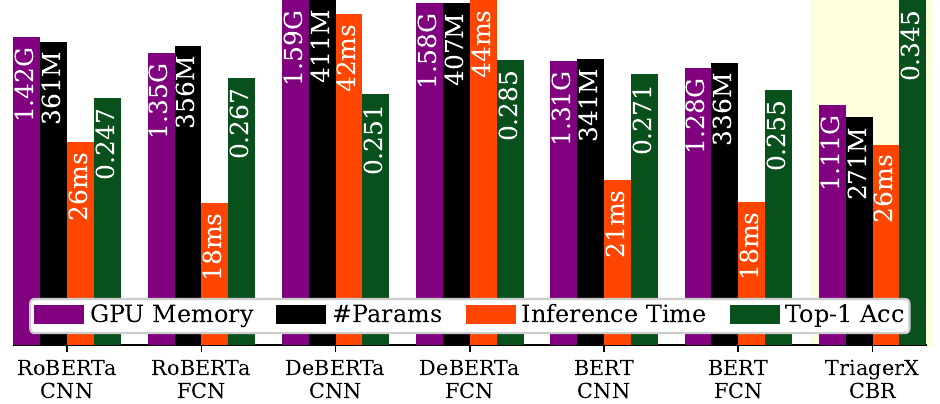}
    \caption{Runtime GPU usage, parameters, inference time, and Top-1 accuracy of CBR vs. baselines on GC dataset.}
    \label{fig:acc-vs-param}
    % \vspace{-1em}
\end{figure}
\rev{Figure~\ref{fig:acc-vs-param} presents a comparison of GPU memory usage, parameter count, inference latency, and Top-1 accuracy in the GC dataset when tested on a NVIDIA V100 GPU. For each model, we compute resource usage by averaging over 50 prediction runs, initializing a fresh runtime each time. CBR achieves higher accuracy despite using fewer parameters than larger PLM-based classifiers. Notably, the closest competitor, BERT-FCN, is approximately 15\% larger in parameter size while delivering lower accuracy. In terms of latency, these models operate in the 18–44 ms range on the GPU, which we consider negligible for real-time usage.} Although CodeBERT-based models offer lower memory consumption due to their smaller size, they consistently underperform in accuracy and are therefore excluded from this plot for clarity. Similarly, we omit LBT-P, a knowledge-distilled model, as TriagerX CBR does not rely on any compression techniques and thus represents a different design trade-off.

%\noindent\textbf{Model Efficiency Needed for Resource Constrained Deployment.}

\vspace{-1em}
\subsection{Component Recommendation in the Client Workflows}
\label{sec:adaptability}
One initial feedback on TriagerX was that recommending component labels is also useful, especially when direct resource allocation is challenging (e.g., developer unavailability). This allows issues to be assigned to the appropriate team for resolution. Hence, we used TriagerX CBR for component classification in Openj9, focusing on all nine key components—\texttt{build, gc, infra, jclextensions, jit, jitserver, jvmti, test, vm}. Each component belongs to a different team. We trained CBR with the same configuration as before (Section \ref{sec:cbr-training}), using issue components as labels. TriagerX CBR outperforms LBT-P by 10.6\% in Top-1 accuracy (Table \ref{table:component-prediction}), along with large PLM-based baselines. We selected RoBERTa and DeBERTa large with CNN classifiers for comparison due to their superior performance in the developer recommendations task among other PLMs. This task benefited from a more balanced data distribution and more samples per class, improving performance compared to developer classification for all models. However, some issues, like issue \href{https://github.com/eclipse-openj9/openj9/issues/19588}{\#19588}, have multiple component labels, impacting Top-1 accuracy, whereas Top-2 and Top-3 accuracies better represent performance in multi-label scenarios.
\begin{table}[t]
\small
\caption{Component prediction accuracy of different models on the OpenJ9 dataset.}
\centering
\resizebox{0.7\linewidth}{!}{
\begin{tabular}{c|ccc}
\toprule
\textbf{Model} & \textbf{Top-1} & \textbf{Top-2} & \textbf{Top-3} \\
\midrule
\textbf{TriagerX CBR} & \textbf{0.782} & \textbf{0.928} & \textbf{0.971} \\
% \midrule
LBT-P & 0.707 & 0.887 & 0.938 \\
% \midrule
RoBERTa-Large + CNN & 0.734 & 0.901 & 0.961  \\
% \midrule
DeBERTa-Large + CNN & 0.745 & 0.902 & 0.957   \\
\bottomrule
\end{tabular}%
}
\label{table:component-prediction}
% \vspace{-1em}
\end{table}

% \vspace{-1em}
\subsection{Usage Feedback from the Industrial Deployment}
\label{sec:deployment-feedback}
%\noindent\textbf{Performance.}
\rev{TriagerX is in early deployment within the OpenJ9 project, using data from the most active contributors, with feedback collection. The model was trained using historical data from 17 active contributors. To assess user adaptation and recommendation quality, we tracked the Top-3 recommendations as per the team's requirement for 80 new issues and measured implicit acceptance based on whether any of them participated in the issue through discussion, pull request, or commit. Using this proxy for acceptance, TriagerX achieved an 82.5\% success rate. We also analyzed a subset of 51 issues that required code contributions to be resolved (e.g., commit). For 34 of these, at least one recommended developer made a code contribution, yielding a 66.7\% acceptance rate.}

% We assessed TriagerX on the OpenJ9 repository by focusing on the actively contributing developers. As per the team's requirements, we assessed the model based on its Top-3 recommendations for each issue. 
% Importantly, this evaluation was conducted on real, newly reported issues that were not part of the model’s training or test set, providing an unbiased measure of generalization in a production environment.
% Within the evaluation window (Issue \#20310 to \#20594), we identified 80 issues where at least one of the 17 developers contributed through assignment, commits, or comments. 
% The model correctly included at least one contributor in its Top-3 predictions for the client at real-time with 82.5\% accuracy. 
% Of these, 51 issues involved actual code contributions (commits or pull requests), while the rest were resolved without code changes (e.g., marked as duplicates). 
% Among the 51 issues with code contributions, the model successfully predicted a contributing developer within its Top-3 recommendations for 34 issues, achieving a 66.7\% acceptance rate in these scenarios.

% Interestingly, in the OpenJ9 repository, developers often report and resolve their own issues. Although the model has no access to reporter identity at inference time, it may still learn associations between issue content and a developer’s historical activity, occasionally resulting in accurate self-recommendations.

%\noindent\textbf{Feedback.} 
The initial reaction from the client team on the use of TriagerX as a triaging assistant is encouraging. TriagerX provided them a valuable starting point, allowing developers to weigh the suggestions against other considerations. One developer remarked, \bf{\textit{``This is a large repo, and I think it’s pretty good what we got—that's a good starting point."}} The developers noted that bug triaging is complex; for example, a second developer may be assigned due to the first's unavailability. \revm{Automated tools may overlook these contextual factors, and the recommendations could be labeled wrong; for example, a developer may be recommended based on technical similarity alone, without accounting for their current availability or workload. These contextual factors represent a direction for improvement, potentially through integrating workload or availability information into the ranking process. Another feature request on TriagerX was to explain the predictions with clear rationales. Developers expressed a desire to understand why a particular developer was recommended, such as knowing that they recently committed to several similar issues, as this would increase their confidence in accepting or overriding the suggestion, representing another direction for improvement we plan to address in future work.}

% Figure \ref{fig:acc-vs-param} compares peak GPU usage, parameter count, inference time, and Top-1 accuracy on the GC dataset. We compute this by averaging 50 prediction instances with fresh runtime each time. TriagerX CBR, using RoBERTa- and DeBERTa-Base is more accurate even with fewer parameters compared to large PLM-based methods with lower GPU usage (the closest PLM-based classifier BERT-FCN is \approx 15\% larger). As for the prediction latency, most of these models perform between 18 - 44 ms which we consider as negligible. While CodeBERT-based models are smaller, resulting in lower memory usage, they perform considerably worse and are excluded from this plot. Our baseline LBT-P, a smaller distilled model, is excluded as TriagerX CBR does not use any model compression technique.

\section{Discussions}
\label{sec:discussion}
% \begin{table}
% \small
% \caption{Top-k accuracy of TriagerX CBR at different contribution threshold.}
% \centering
% \resizebox{6cm}{!}{
% \begin{tabular}{c|cccc}
% \toprule
% \textbf{Dataset} & \textbf{Threshold} & \textbf{\#Developers} & \textbf{Top-1} & \textbf{Top-3} \\
% \midrule
% \multirow{5}{1.5cm}{\centering Google Chromium}  & 5 & 1641 & 0.326 & 0.505\\
% & 10 & 1320 & 0.329 & 0.513\\
% & 15 & 1120 & 0.330 & 0.516 \\
% & 20 & 1032 & 0.345 & 0.537 \\
% & 25 & 886 & 0.345 & 0.546\\
% \midrule
% \multirow{5}{1.5cm}{\centering Openj9} & 5 & 103 & 0.232 & 0.423 \\
% & 10 & 72 & 0.246 & 0.442 \\
% & 15 & 57 & 0.260 & 0.471\\
% & 20 & 50 & 0.272 & 0.476 \\
% & 25 & 44 & 0.281 & 0.486 \\
% \bottomrule
% \end{tabular}%
% }
% \label{table:contribution-threshold}
% \vspace{-1em}
% \end{table}

\begin{table}
\small
\caption{Top-k accuracy of TriagerX CBR and other baselines at different contribution thresholds.}
\centering
\resizebox{\linewidth}{!}{
\begin{tabular}{c|c|c|cc|cc|cc}
\toprule
\multirow{2}{*}{\centering\textbf{Dataset}} & \multirow{2}{*}{\textbf{Threshold}} & \multirow{2}{*}{\textbf{\#Dev}} &
\multicolumn{2}{c|}{\textbf{TriagerX CBR}} &
\multicolumn{2}{c|}{\textbf{\revtwo{LBT-P}}} & \multicolumn{2}{c}{\textbf{\revtwo{RoBERTa-L (CNN)}}} \\
\cmidrule(lr){4-5} \cmidrule(lr){6-7} \cmidrule(lr){8-9}
 & & & \textbf{Top-1} & \textbf{Top-3} & \textbf{\revtwo{Top-1}} & \textbf{\revtwo{Top-3}} & \textbf{\revtwo{Top-1}} & \textbf{\revtwo{Top-3}} \\
\midrule
\multirow{5}{1.5cm}{\centering Google Chromium}  
& 5  & 1641 & 0.326 & 0.505 & \revtwo{0.296} & \revtwo{0.459} & \revtwo{0.251} & \revtwo{0.425} \\
& 10 & 1320 & 0.329 & 0.513 & \revtwo{0.309} & \revtwo{0.484} &  \revtwo{0.262} & \revtwo{0.438} \\
& 15 & 1120 & 0.330 & 0.516 & \revtwo{0.305} & \revtwo{0.487} & \revtwo{0.268} & \revtwo{0.449} \\
& 20 & 1032 & 0.345 & 0.537 & \revtwo{0.318} & \revtwo{0.499} & \revtwo{0.281} & \revtwo{0.475} \\
& 25 &  886 & 0.345 & 0.546 & \revtwo{0.327} & \revtwo{0.513} & \revtwo{0.289} & \revtwo{0.488} \\
\midrule
\multirow{5}{1.5cm}{\centering Openj9} 
& 5  & 103 & 0.232 & 0.423 & \revtwo{0.181} & \revtwo{0.304} & \revtwo{0.131} & \revtwo{0.264} \\
& 10 &  72 & 0.246 & 0.442 & \revtwo{0.188} & \revtwo{0.319} & \revtwo{0.165} & \revtwo{0.330} \\
& 15 &  57 & 0.260 & 0.471 & \revtwo{0.200} & \revtwo{0.405} &  \revtwo{0.185} & \revtwo{0.368} \\
& 20 &  50 & 0.272 & 0.476 & \revtwo{0.211} & \revtwo{0.407} & \revtwo{0.206} & \revtwo{0.403} \\
& 25 &  44 & 0.281 & 0.486 & \revtwo{0.230} & \revtwo{0.412} & \revtwo{0.209} & \revtwo{0.411} \\
\bottomrule
\end{tabular}%
}
\label{table:contribution-threshold}
% \vspace{-1em}
\end{table}

% While TriagerX improves performance over the baselines, it still has some limitations. We discuss some limitations in this section.

\noindent\textbf{Impact of Data Imbalance.} 
\rev{The data imbalance problem is common across the SOTA bug triaging techniques we studied. In real-world repositories, a small group of \textit{star developers} resolve the majority of issues, causing ML/DL models to disproportionately favor these frequent developers, reducing their effectiveness in recommending less active contributors. 
}

\revtwo{Mani et al.~\cite{mani2018deeptriage} first showed that DL-based bug triaging models gain from increasing the number of training samples per developer and improvements plateau around a threshold of 20, particularly on large datasets like GC. To further examine how different approaches respond to class imbalance, we compare TriagerX CBR with two strong baselines--LBT-P and RoBERTa-Large (CNN)--under varying contribution thresholds on two datasets of very different sizes: GC and Openj9. As shown in Table~\ref{table:contribution-threshold}, increasing the minimum contribution threshold (i.e., filtering out infrequent developers) improves accuracy across all models. However, TriagerX CBR consistently outperforms both baselines at every threshold. Finally, we use a threshold of 20 in most experiments, as it offers a good balance between predictive accuracy and developer coverage, and is consistent with prior findings.}

\rev{While contribution thresholding improves predictive accuracy by reducing the impact of imbalance, it also reduces the number of developers considered, potentially excluding less experienced or newer contributors. To mitigate imbalance without aggressively narrowing coverage, we evaluated alternative techniques including oversampling, undersampling, SMOTE~\cite{chawla2002smote}, focal loss~\cite{lin2017focal}, and weighted cross-entropy. Most of these approaches either degraded performance or destabilized training; weighted sampling proved the most effective and is therefore integrated into our final pipeline.}

\revth{Importantly, while contribution thresholding reduces the size of the developer set, it does not substantially reduce issue coverage compared to using no threshold at all. For example, on the Google Chromium dataset, increasing the contribution threshold from 5 to 20 retains 62.8\% of developers (Table~\ref{table:contribution-threshold}), yet these developers still account for 91.8\% of all issues relative to no thresholding. A similar trend holds for OpenJ9, where 82.0\% of issues remain covered at the same threshold. This shows that a relatively small group of consistently active developers resolves most issues, and that the chosen threshold reflects a practical trade-off between coverage and reliability rather than an arbitrary exclusion.}

% \noindent\textbf{IBR's Performance on Sparse Interaction.} TriagerX IBR relies on historical developer interactions, such as commits and comments, to rank developers. This reliance can limit its effectiveness in projects with sparse activity or for developers with minimal contribution history. TriagerX framework addresses this challenge to some extent by allowing the influence of IBR to be tuned based on repository characteristics. 
% Moreover, when interaction data is absent (e.g., literature datasets used in this paper), the CBR component remains effective by leveraging the semantics of issue text.

\noindent\textbf{Cold Start.}  \revtwo{Cold-start is another fundamental challenge for ML-based bug triaging approaches, as developers without prior history (e.g., new joiners) cannot be effectively modeled. In TriagerX, CBR may struggle to assign bugs to infrequent contributors when labeled training data is limited. Similarly, IBR relies on past interactions (e.g., commits) to build developer representations, which can make ranking developers with sparse/no activity more difficult.} 

\revtwo{TriagerX mitigates this challenge in two ways. First, the influence of IBR ($W_f$ in Equation~\ref{eq:rank-agg}) can be tuned based on repository characteristics, offering flexibility for projects with sparse developer histories. Second, in collaboration with our industry partner, we extended TriagerX to also predict the component associated with an issue. Since each component is owned by a team, this recommendation functions as a team-level assignment. The corresponding team lead can then assign the issue to a suitable developer, including new joiners. This setup ensures continuity despite developer turnover and allows new developers to gradually build a contribution history. Over time, these contributions can be incorporated into TriagerX retraining, enabling automated recommendations. While this approach does not eliminate the cold-start problem—an inherent limitation of all learning-based triaging models—it provides a practical and effective mechanism to partially mitigate it in industrial settings.} \revm{For instance, consider a scenario where a new developer joins the OpenJ9 component team responsible for \texttt{comp:jit}. Without sufficient contribution history, TriagerX's CBR and IBR components cannot directly recommend them. However, TriagerX's component-level recommendation can correctly route the issue to the \texttt{jit} team, allowing the team lead to assign it to the new developer. Over time, as the developer resolves issues and builds a contribution history, TriagerX can incorporate their activity in subsequent retraining cycles, gradually enabling direct recommendations. Looking further ahead, the cold-start problem could be more directly addressed by leveraging pre-trained developer profiles derived from public contribution histories on platforms like GitHub, enabling the model to bootstrap representations for new developers before they accumulate sufficient local history. Similarly, transfer learning techniques could be used to adapt developer representations from repositories with rich interaction data to those with sparse histories, reducing the cold-start window for new contributors.}

\noindent\textbf{Threats to Validity.} The \textbf{internal validity} of this study may be impacted by biases in dataset selection and preprocessing. Using different training and testing samples than LBT-P was necessary to capture our datasets' unique characteristics but introduced discrepancies. Despite standardizing preprocessing and hyperparameters, variations in dataset handling could still lead to confounding factors. Additionally, reproducing baselines like LBT-P and MDN may result in inconsistencies, even when following the authors' guidelines and the original paper.
The \textbf{generalizability} of our findings may be constrained by the specific datasets and projects used in this study. \rev{Although we evaluated TriagerX across multiple, diverse repositories, its performance may not directly transfer to unseen projects with different developer populations, contribution patterns, or codebase characteristics. In particular, models trained on one repository are unlikely to perform well on another due to mismatches in label spaces and developer behavior.} \revm{Furthermore, the validity of our industrial deployment evaluation is subject to limitations in how acceptance is measured. The acceptance rates reported in Section \ref{sec:deployment-feedback} reflect post-triage developer activity rather than direct triager acceptance, as developer engagement with an issue does not necessarily indicate that the triager was explicitly assigned based on TriagerX's recommendation. Unobserved confounders such as developer availability, current workload, or manual override decisions may independently influence participation regardless of the recommendation. Therefore, these metrics serve as a behavioral proxy for acceptance rather than a precise measure of recommendation adoption.}

%\input{sections/threats}
% \vspace{-1em}
\section{Related Work}
Broadly, automated bug triaging techniques are two types: (1) content-based and (2) developer activity-based \cite{sarkar2019ericsson, guo2020devactivity}.

% Bug triage has been extensively studied within the research community. Broadly, bug triaging techniques can be categorized into two types: content-based and developer activity-based \cite{sarkar2019ericsson, guo2020devactivity}. Content-based bug triaging primarily focuses on the textual content of the bug report, leveraging natural language processing and information retrieval techniques to match bugs with appropriate developers. Developer activity-based approaches consider the historical activity or interactions and expertise of developers, using their past contributions and interactions to inform triaging decisions.

\noindent\textbf{Content-based.} Numerous approaches consider bug triaging as a classification problem \cite{Cubranic2004AutomaticBT, mani2018deeptriage, lee2023lbtp, lee2017dlbasedtriager, wei2020efficientbugtriage} and exploit algorithms like Na\"{i}ve Bayes and SVM with TF-IDF \cite{Cubranic2004AutomaticBT, anvik2006svm, anvik2011reducing, ahsan2009svm, Kanwal2012BugPT, dedik2016automated, fu2017easyoverhard}. Ensemble learning was also evaluated \cite{johnsson2016ensemble}. Many also approached the problem through information retrieval (IR) \cite{linares2012triaging, Shokripour2013complicated}. Recent studies mostly focus on using deep learning techniques like bi-directional RNNs \cite{mani2018deeptriage} with Word2Vec, CNNs with both context-free \cite{lee2017dlbasedtriager, mani2018deeptriage} and context-sensitive \cite{zaidi2020cnntriage, lee2023lbtp} word embedding techniques like ELMo \cite{peters-etal-2018-deep} and RoBERTa, and hierarchical attention \cite{he2021hierarchicalattention}. Dual-output networks are also used to predict team and developer assignments \cite{Choquette2019dualdnn}. Dipongkor et al. \cite{dipongkar2023comparison} showed that different PLMs exhibit substantial orthogonality in bug triaging and suggested improvements through PLM ensembles. \textit{Our study differs on usage of bug representations via an ensemble of PLMs that is smaller than SOTA PLMs but achieves better result than those.}

\noindent\textbf{Developer Activity-based.} Leveraging developer expertise scores can also improve bug triaging \cite{Tamrawi2011fuzzy, hu2014effective, Tian2016rank}.  Methods include topic modeling as Latent Dirichlet Allocation (LDA) \cite{blei2003lda,xie2012dretom, zhang2014developerranking, yang2014multimodalbugtriaging, xia2017lda} and socio-technical models for prioritizing developers based on contributions \cite{xuan2012prioritization}. However, these methods may struggle with ambiguous language and require extensive annotated data \cite{zhang2013heterogeneous}. ML techniques using tossing graphs use developers' tossing history \cite{jeong2009tossing, Bhattacharya2010toss, Xi2019BugTB, su2021tossing}. IR combined with ML and DL explores developer recommendations based on code vocabulary \cite{matter2009vocab}, interaction metrics \cite{yang2014recommendation}, and activity profiles \cite{naguib2013activity}. Hybrid algorithms and neural networks, including CNNs \cite{guo2020devactivity} and ensemble approaches \cite{zhang2022sustriage} that groups developers based on their activity, improve accuracy but these approaches also face limitations, such as reliance on specific vocabulary \cite{matter2009vocab} or insufficient consideration of contextual information in bug reports \cite{guo2020devactivity}. \textit{We differ in the types of interactions used and their scoring methods and by enriching the final recommendation with both content and interaction based rankings.}

\section{Conclusion \& Future Work}
In this paper, we have presented TriagerX, a SOTA bug triaging framework that consistently outperforms recent baselines across all literature and industrial partner datasets. We do this by ensembling two novel ranking models: Content-Based Ranker (CBR) and Developer Interaction Based Ranker (IBR). 
% The CBR utilizes multiple smaller PLMs to analyze textual aspects, outperforming larger PLMs and existing SOTA baselines. The IBR leverages historical interaction data to rank developers based on expertise. Together, these components enhance the relevance and effectiveness of bug triage recommendations. 
Future work will focus on (1) developing a feedback-enhanced auto-adaptable bug triaging model that could learn from user feedback and improve its triaging recommendations, (2) offering a human-centric approach to use TriagerX by offering contextual explanations of recommendations using IBR's internal signals (e.g., number of commits made by a developer on similar issues) in combination with LLM-based reasoning — such explanations could increase the confidence of human triagers and improve trust in automated decisions, and \revm{(3) extending TriagerX to related software engineering tasks such as code review and task assignment, where matching contributors to work items based on both textual content and historical activity patterns can be critical. Additionally, incorporating richer contextual signals, such as developer collaboration networks, workload history, and team structures, could further improve recommendation accuracy, particularly for repositories with complex contributor dynamics or frequent developer turnover.}

% \section{Data Availability}
% All source codes including the datasets are available here: \url{https://anonymous.4open.science/r/triagerX-37AF}

\appendices

\section{PLM Variants}
\label{app:plm-variants}

\begin{table}[t]
\small
\caption{Overview of different PLMs.}
\centering
\resizebox{\linewidth}{!}{
\begin{tabular}{c|cccc}
\toprule
\textbf{PLM} & \textbf{\#Params} & \textbf{Hidden Size} & \textbf{Layers} & \textbf{Attention Heads} \\
\midrule
BERT-Base    & 110M  & 768  & 12 & 12 \\
BERT-Large   & 335M  & 1024 & 24 & 16 \\
RoBERTa-Base & 125M  & 768  & 12 & 12 \\
RoBERTa-Large& 355M  & 1024 & 24 & 16 \\
DeBERTa-Base & 139M  & 768  & 12 & 12 \\
DeBERTa-Large& 405M  & 1024 & 24 & 16 \\
CodeBERT     & 125M  & 768  & 12 & 12 \\
\bottomrule
\end{tabular}%
}
\label{table:plm-variants-comparison}
% \vspace{-1em}
\end{table}

Pretrained Language Models (PLMs) are available in different sizes, offering trade-offs between performance, efficiency, and deployment requirements. Larger variants capture richer representations but require more memory and computation, making model selection dependent on task-specific needs.
Table \ref{table:plm-variants-comparison} provides an overview of the PLM variants used in this study, highlighting key architectural differences. All of the models were sourced and evaluated from the HuggingFace repository \cite{huggingface}.

\section{Additional Baselines}
\label{app:additional-baselines}
We provide reproduction results for two recent graph-based bug triaging methods, NCGBT \cite{dong2024neighborhood} and PCG \cite{dai2024pcg}, which are omitted from the main manuscript due to their inferior performance on our datasets. Both methods were reproduced using their official implementations. NCGBT performed poorly across all datasets, with performance degrading sharply as the dataset size increased. This is likely due to its reliance on CC fields, which are absent in our datasets. PCG does not depend on CC data but exhibits very low Top-k accuracy on large projects, consistent with what the original PCG paper reported for Google Chromium. Given PCG's near-zero accuracy on the large-scale GC dataset, we did not extend its evaluation to the remaining datasets, as meaningful results were unlikely. The complete results are shown in Table~\ref{table:additional-baseline-reproduction}. We provide detailed reproduction steps on these baselines in our replication package\footnote{\url{https://github.com/afifaniks/triagerX}}.

\begin{table}[t]
\small
\caption{Accuracy of different graph-based approaches in our datasets.}
\centering
\resizebox{\linewidth}{!}{
\begin{tabular}{llccccc}
\toprule
\textbf{Method} & \textbf{Dataset} & \textbf{Top-1} & \textbf{Top-3} & \textbf{Top-5} & \textbf{Top-10} & \textbf{Top-20} \\
\midrule
\multirow{3}{1.5cm}{NCGBT} 
& GC & 0.0036 & 0.0098 & 0.0148 & 0.0264 & 0.0477 \\
& OpenJ9 & 0.0353 & 0.1191 & 0.1700 & 0.2848 & 0.4437 \\
& TS & 0.0503 & 0.1661 & 0.2648 & 0.4255 & 0.7065 \\
\midrule
PCG & GC & 0.0076 & 0.0102 & 0.0165 & 0.0293 & 0.0560 \\
\bottomrule
\end{tabular}
}
\label{table:additional-baseline-reproduction}
% \vspace{-1em}
\end{table}

\balance
\bibliographystyle{IEEEtran}
\bibliography{references}

% \newpage
\end{document}